\documentclass[12pt,preprint]{aastex}
\usepackage{epsfig}
\usepackage{psfig}

\newcommand{\ASCA}{{\sl ASCA}}

\newcommand{\Chandra}{{\sl Chandra}}
\newcommand{\ROSAT}{{\sl ROSAT}}

\newcommand{\et}{{\it et~al.}}

\newcommand{\EG}{3EG~J2020+4017}
\newcommand{\RX}{RX~J2020.2+4026}
\newcommand{\CG}{2CG078+2}

\newcommand{\gray}{{$\gamma$-ray}}
\newcommand{\grays}{{$\gamma$~rays}}

\slugcomment{\apj, submitted 2006.06.22}
\shorttitle{{A search for a counterpart of the unidentified $\gamma$-ray source 3EG J2020+4017 (2CG078+2)}}
\shortauthors{W.~Becker et al.}

\begin{document}

\title{{A search for the X-ray counterpart of the unidentified \gray\ source \EG\ (2CG078+2)}}

\author{
Martin~C.~Weisskopf\altaffilmark{1},
Douglas~A.~Swartz\altaffilmark{2},
Alberto~Carrami\~nana\altaffilmark{3,4},
Luis~Carrasco\altaffilmark{3,4},
David~L.~Kaplan\altaffilmark{5},
Werner~Becker\altaffilmark{6},
Ronald~F.~Elsner\altaffilmark{1},
Gottfried~Kanbach\altaffilmark{6},
Stephen~L.~O'Dell\altaffilmark{1},
Allyn~F.~Tennant\altaffilmark{1}
}

\altaffiltext{1}
{NASA Marshall Space Flight Center, Space Science Office, VP62, Huntsville, AL 35812 USA.}
\altaffiltext{2}
{Universities Space Research Association, NASA Marshall Space Flight Center, Space Science Office, VP62, Huntsville, AL 35812 USA.}
\altaffiltext{3}
{Instituto Nacional de Astrof\'{\i}sica, \'{O}ptica y Electr\'{o}nica, Luis Enrique Erro 1, Tonantzintla, Puebla 72840, M\'{e}xico.}
\altaffiltext{4}
{Visiting Astronomer, Observatorio Astrof\'{\i}sico Guillermo Haro, Cananea, Sonora, M\'{e}xico.}
\altaffiltext{5}
{Pappalardo Fellow; Kavli Institute for Astrophysics and Space Research, Massachusetts Institute of Technology, 37-664H, 77 Massachusetts Avenue, Cambridge, MA 02139 USA.}
\altaffiltext{6}
{Max-Planck-Institut f\"{u}r extraterrestrische Physik, 85741 Garching bei M\"{u}nchen, Germany.}

\begin{abstract}
We report observations with the {\sl Chandra X-ray Observatory} of a field in the $\gamma$-Cygni supernova remnant (SNR78.2+2.1) centered on the cataloged location of the unidentified, bright \gray\ source \EG.
In this search for an X-ray counterpart to the \gray\ source, we detected 30 X-ray sources.
Of these, we found 17 strong-candidate counterparts in optical (visible through near-infrared) cataloged and an additional 3 through our optical observations.
Based upon colors and (for several objects) optical spectra, nearly all the optically identified objects appear to be reddened main-sequence stars:
None of the X-ray sources with an optical counterpart is a plausible X-ray counterpart to \EG---if that \gray\ source is a spin-powered pulsar.
Many of the 10 X-ray sources lacking optical counterparts are likely (extragalactic) active galactic nuclei, based upon the sky density of such sources.
Although one of the 10 optically unidentified X-ray sources could be the \gray\ source, there is no auxiliary evidence supporting such an identification.
\end{abstract}


\section{Introduction\label{s:intro}}

\EG\ is the brightest of the many unidentified sources detected by the Energetic Gamma-Ray Experiment Telescope (EGRET).
The third and final official EGRET catalog 3EG \citep{har99} lists 271 sources, of which only a quarter have firm identifications.
The paucity of identifications results primarily from large uncertainties in \gray-source positions, due to the angular resolution of EGRET's multilevel-spark-chamber detector and to poor photon statistics.
Typical positional uncertainties are a few degrees \citep{mat96}, improving to $\approx\!0.5\arcdeg$ (FWHM) for bright sources with adequate statistics above several GeV, for which determination of the photon's direction is more precise.
Systematic uncertainties---due to non-uniform background or the presence of adjacent sources---also can affect positional determinations \citep{har99}.
This is most severe for sources near the Galactic plane, where the diffuse background is bright and highly structured and the concentration of \gray\ sources is high.
Consequently, most EGRET sources identified through positional correlations are  extragalactic---e.g., blazars---at high Galactic latitude.
Even so, about half the high-confidence blazar identifications from the 3EG catalog lie outside the stated 68\%-confidence contours \citep{har99}.

Clearly, identifying counterparts to \gray\ sources at low Galactic latitude requires more than positional correlations.
For example, the identification of seven EGRET sources as spin-powered pulsars relied upon detection of periodic modulation of the \gray\ flux.
Based upon pulsar population models \citep{mcl00,zha00}, pulsar emission models \citep{rom95,gon04}, the association of EGRET sources with young objects, and the fact that the only definite identifications of persistent EGRET Milky-Way sources are bright pulsars, we expect many more EGRET sources to be pulsars.
However, further searches for pulsations in EGRET data---either blind \citep{cha01} or folding on known pulsar periods \citep{nel96}---have been unsuccessful.

Analyses of the sample distribution of \gray\ spectrum, flux, variability, and location has produced some progress in distinguishing classes of EGRET sources.
A population of hard, steady, Galactic-plane sources seem similar to confirmed \gray\ pulsars; another of soft, variable, high-Galactic-latitude sources are likely unidentified blazars \citep{sow04}.
In addition, there are a group of weaker, steady sources, spatially correlated with the Gould belt \citep{geh00}; a group of luminous, soft, highly variable (and perhaps older) sources that form a halo about the Galactic center \citep{gre01,gre04}; and a subgroup of variable Galactic-plane objects \citep{nol03}.
While the nature of most of these \gray\ sources is unknown, spatial correlations with tracers of recent star formation---including supernova remnants \citep[SNRs;][]{stu95}, OB associations \citep{kaa96}, and massive stars \citep{rom99}---support an association with young objects.

Proposed \gray-emission mechanisms are consistent with young source populations.
Shocks in massive-star supersonic winds \citep{cas80}, high-mass X-ray binaries \citep{tav97,kir99}, microquasar jets \citep{geo02}, pulsar magnetospheres, and pulsar-wind nebulae \citep{rob05} are all environments conducive to producing high-energy \grays---either directly through electron bremsstrahlung, Compton scattering, or synchrotron emission; or indirectly through hadronic collisions with surrounding material followed by $\pi^o$-decay emission.
Such processes also work on larger spatial scales: Cosmic rays produced in supernova blast waves can interact with the interstellar medium to produce \grays\ \citep{gai98}.
These same mechanisms may produce the diffuse \gray\ background \citep{ber93,hun97}.

\EG\ is in the Galactic plane \citep[$l,b=78.05,2.08$;][]{har99}, as are all the known \gray\ pulsars.
It exhibits a steady \citep{nol03}, hard \citep[photon index $\Gamma=2.08\pm0.04$;][]{esp96}, power-law \gray\ spectrum breaking at about 4\,GeV \citep{mer96,buc98}, typical of spin-powered pulsars \citep{tho99}.
Hence, \EG\ is one of the best candidates for a \gray\ pulsar \citep{mer96}.
However, efforts to detect pulsations in EGRET data \citep{bra96,koh95,cha01} or to discover radio pulsars in the region \citep{bec04,nic97} have been unsuccessful.
Pulsations are difficult to detect in EGRET data, due to low signal-to-noise ratios and to the propensity of \gray\ pulsars to spin-down rapidly and to exhibit glitches \citep[e.g.,][]{cha01}.
In fact, the successful detection of \gray\ pulsations in most EGRET pulsars resulted from epoch-folding the \gray\ data with the known period of the pulsar \citep{nol93,kan94,ram95}.

The absence of a radio source---to a sensitivity of about 40\,$\mu$Jy \citep{bec04}---within the \EG\ error circle, is not unprecedented amongst \gray\ pulsars:
The nearby Geminga pulsar is radio quiet.
The discovery of pulsations in a coincident \ROSAT\ X-ray source \citep{hal92}, followed by verification in the EGRET data \citep{ber92}, are the basis for Geminga's classification as a pulsar.
With Geminga as an archetype, several authors \citep[e.g.,][]{hel94,der94,tho94,rom95} have suggested that radio-quiet \gray\ pulsars may account for most of the unidentified EGRET sources.
Likewise, the recently discovered \citep{obr05} transient radio pulsars---Repeating Radio Transients (RRATs) and intermittent (short-burst-mode) radio pulsars---may also contribute to this population.

If \EG\ is a radio-quiet pulsar similar to Geminga, then we can estimate its X-ray emission, under certain assumptions.
A 0.5-MK blackbody component dominates Geminga's soft-X-ray flux \citep{kar05,hal97}, producing a ratio of energy flux $S_{x}/S_{\gamma}=1.8\!\times\!10^{-3}$ for the 0.1--2.4-keV and 0.1-5-GeV bands \citep{tho96}.
From \ROSAT\ observations, we \citep{bec04} previously set an upper limit $S_{x}<1.8\!\times\!10^{-13}$~erg/(cm$^{2}$ s) to any soft, point-like X-ray counterpart to \EG:
This corresponds to $S_{x}/S_{\gamma}\leq 2.3\!\times\!10^{-4}$, well below expectations based on Geminga.
The analog of the harder, nonthermal, X-ray emission from Geminga is better matched to the effective bandpass of the \Chandra\ ACIS I array.
\citet{jac02} measure a 0.7--5.0-keV flux $2.6\!\times\!10^{-13}$~erg/(cm$^{2}$ s) from Geminga. 
Again, if we assume a similar $S_{x}/S_{\gamma}$ ratio, we would expect a flux $<10^{-13}$~erg/(cm$^{2}$ s) from \EG.
Our \Chandra\ observation easily detects X-ray sources at this level; however, there are dozens of faint X-ray sources within the error contour for \EG.

The direction of \EG\ lies within the 1\arcdeg-diameter \citep{hig77}, young supernova remnant SNR78.2+2.1 \citep{gre74}, often called the ``$\gamma$-Cygni SNR'' after the unrelated bright near-by star.
The distance to the $\gamma$-Cyg SNR is approximately 1.6~kpc \citep{hig77,lan80}, corresponding to a (true) distance modulus $\Delta m \approx 11.0$.

Having noticed the proximity of the \gray-source and SNR directions, \citet{stu95} suggested that the \gray\ emission from \EG\ could result from protons accelerated in the interaction of the supernova blastwave with surrounding gas.
They emphasized that the SNR's young age, radio brightness, and flat radio spectrum all favored this cosmic-ray acceleration mechanism.
Following \citet{dru94}, they also remarked that the higher density medium of the molecular cloud associated with the SNR should enhance the \gray\ flux.

Using similar arguments, \citet{esp96} concluded the $\gamma$-Cyg SNR is one of the best candidate cosmic-ray origin sites.
However, they found that the \gray\ spectrum of \EG\ is much flatter than expected from $\pi^{\rm o}$ decay.
\citet{gai98} showed that larger contributions from electron bremsstrahlung and Compton scattering would make the model more consistent with observations.
However, the model is still inconsistent with observed upper limits to the TeV \gray\ emission from \EG\ \citep{buc98}, thus questioning its applicability to this source.
Furthermore, the EGRET data are more consistent with a point source than with the extended SNR emission.

X-ray searches for cosmic-ray acceleration sites in the $\gamma$-Cyg SNR have been unsuccessful.
While \ROSAT\ \citep{bra96,bec04}, \ASCA\ \citep{uch02}, and {\sl INTEGRAL} \citep{byk04} observations all detect X-ray emission within the remnant, none originates near the cataloged position of \EG.
The soft X-ray emission could be obscured by intervening material.
However, contrary to expectations for the cosmic-ray acceleration model \citep[e.g.,][]{byk00}, there is no radio emission from the location of \EG.

\citet{bra96} analyzed six \ROSAT\ PSPC observations of the $\gamma$-Cyg SNR region, pointed within 40\arcmin\ of the reported position of \CG.
They found a point source---\RX---within the 95\%-confidence 
contour of the 2EG position, which they suggested as the X-ray counterpart to the \gray\ source.
Further, \citet{bra96} and \citet{car00} found a possible optical counterpart for \RX.
Follow-up optical observations identified a $14.5^{\rm mag}$ K0-V star within the $\approx\!6\arcsec$-radius \ROSAT\ error circle, which would not account for the \gray\ emission.
In that the X-ray--to--optical flux ratio is only marginally consistent with that of late-type stars \citep{sto91,fle95,bra96} one could not totally exclude the association of \RX\ with the \gray\ source.

The 3EG catalog \citep{har99} gave an improved  position for \EG---$20^{\rm h} 21^{\rm m} 1\fs0 +40\arcdeg 17\arcmin 48\arcsec$---displaced by about 10\arcmin\ from the 2EG position used by Brazier \et\ (1996).
With this improved position, the proposed counterpart \RX\ no longer lies within the 95\%-confidence contour of \EG, although it is within the 99\%-confidence contour. 
Previously \citep{bec04}, we reported follow-up studies of \RX\ with \Chandra\ and the Green Bank Radio Telescope. 
The \Chandra\ observations determined the position and spectrum of \RX\ with high precision, showing that \RX\ is indeed the X-ray counterpart to the K0-V star, with a soft spectrum indicative of stellar coronal emission.
The Green Bank observations set upper limits to the presence of radio pulsations.

Here we report results of a \Chandra\ observation centered on the 3EG position for \EG.
The \Chandra\ sensitivity is so deep that many X-ray sources lie within the formal error contours for the position of \EG.
Consequently, we seek a distinguishing feature amongst the detected X-ray sources, in order to establish one (or more) of them as a probable counterpart to \EG.
Section~\ref{s:xray_obs} describes the results of our search for point-like X-ray sources.
Section~\ref{s:oi} presents the results of our searches for candidate optical counterparts to the detected X-ray sources through cross correlating with visible and near-infrared catalogs, and through new optical imaging and spectroscopic observations.
Finally, Section~\ref{s:summary} discusses and summarizes the results.

\section{\Chandra\ Observations and Data Analysis \label{s:xray_obs}}

We obtained a 14.3-ks \Chandra\ observation (ObsID 5533, 2005 February 6) using the Advanced CCD Imaging Spectrometer (ACIS) imaging (I) array (CCDs I0,1,2,3) in the faint, timed-exposure mode, with 3.141-s frame time.
Background levels were nominal throughout the observation. 
Standard \Chandra\ X-ray Center (CXC) processing (version DS.7.4) provided accurate aspect determination.
Starting with level-1 event lists, we cleaned the data column-by-column and de-randomized event positions in order to improve the on-axis point spread function (PSF), thus enhancing the source-detection efficiency and positional accuracy.
In searching for sources, we utilized events in pulse-invariant channels corresponding to 0.511 to 8.030 keV.

\begin{figure}
\epsfig{file=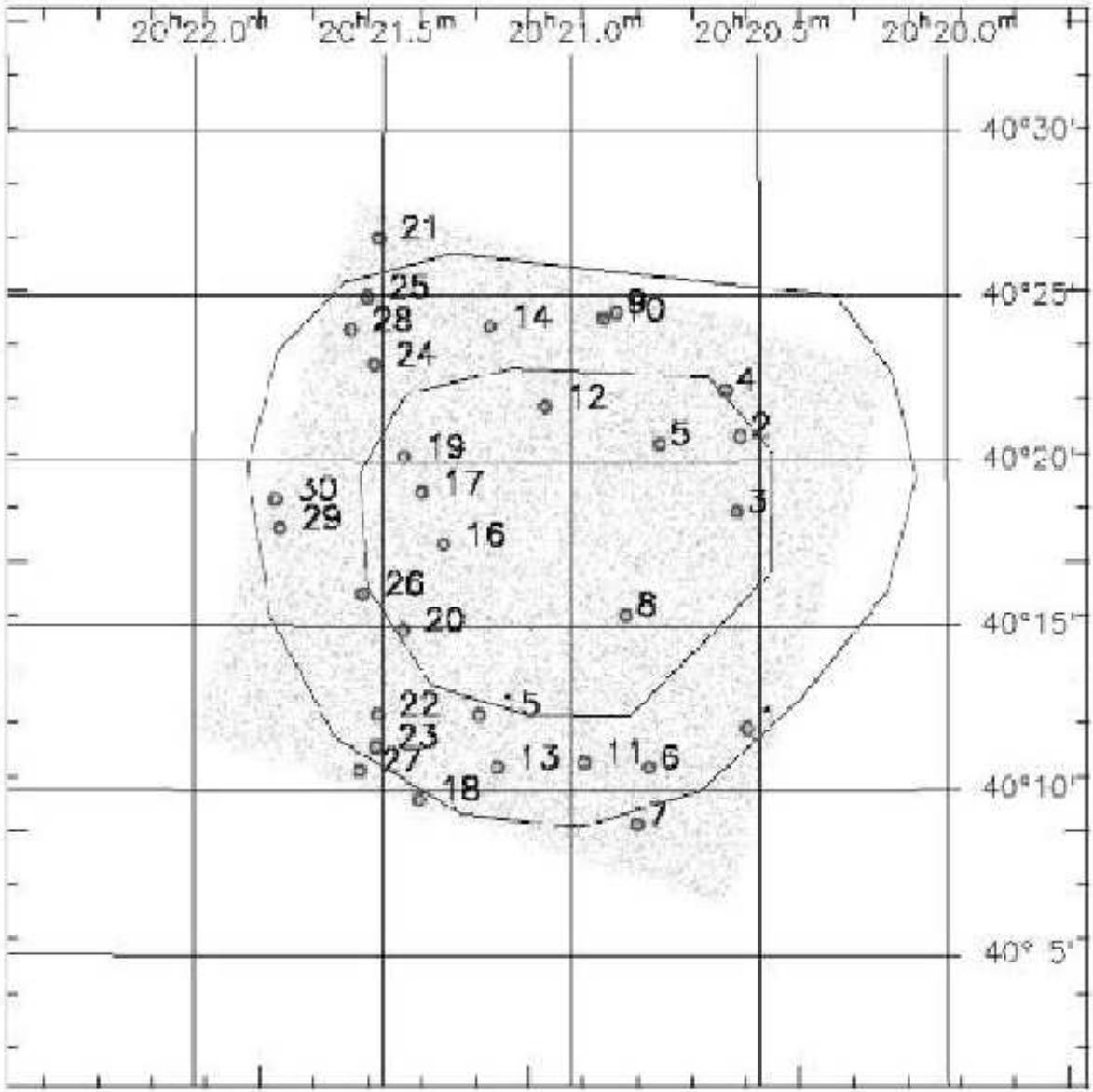,width=16cm,angle=0}
\figcaption{{\Chandra\ image of the \EG\ field.
An open circle encloses each of the 30 detected X-ray sources.
Contours delimit 68\%- and 95\%-confidence levels for the position of \EG.}
\label{f:chandra}}
\end{figure}

The \Chandra\ pointing---${\rm RA(J2000)} = 20^{\rm h}\,21^{\rm m}\,2\fs44$ and Dec(J2000) = 40\degr 17\arcmin 46\arcsec---was only 17\arcsec\ from the best EGRET position (\S\ref{s:intro}).
Figure~\ref{f:chandra} shows the ACIS-I image, overlaid with the 3EG likelihood contours for the cataloged location of \EG\ and with a small circle at the position of each \Chandra-detected source (\S\ref{s:analysis_image}).
Section~\ref{s:analysis_image} describes the analysis of the X-ray imaging; 
Section~\ref{s:analysis_spec}, of the X-ray spectra.
Based upon the measured count rates and an assumed spectral form, Section~\ref{s:analysis_flux} estimates the X-ray flux of the detected sources.
Section~\ref{s:analysis_var} presents indications of X-ray variability in two sources.

\subsection{Image Analysis \label{s:analysis_image}}

We searched for X-ray sources employing techniques described in \citet{ten06}, using a circular-Gaussian approximation to the point spread function (PSF) and setting the signal-to-noise (S/N) threshold for detection to 2.4.
The resulting background-subtracted point-source detection limit is about 7 counts, with fewer than 1 accidental detection expected over the field.
Based upon tests on |chandra\ deep fields, this approach finds all X-ray sources in \Chandra\ fields down to 10 counts, which we thus regard as the completeness limit.

Table~\ref{t:data_x} tabulates  X-ray properties of the 30 {\sl Chandra}-detected sources, denoted in column~1 as S$s$ with $s=\{01, 30\}$.
Columns~2--5 give, respectively, right ascension RA, declination Dec, extraction radius $\theta_{\rm ext}$, and approximate number of X-ray counts $C_{x}$ detected from the source.
Column~6 lists the single-axis RMS error $\sigma_{x}=[(\sigma_{\rm PSF}^2/C_{x}) + \sigma^2_{\rm sys}]^{1/2}$ in the X-ray-source position, where $\sigma_{\rm PSF}$ is the dispersion of the circular Gaussian that approximately matches the PSF at the source location and $\sigma_{\rm sys}$ is a systematic error.
Uncertainties in the plate scale\footnote{See
http://asc.harvard.edu/cal/hrma/optaxis/platescale/} imply $\sigma_{\rm sys}\approx 0\farcs13$: 
To be conservative, we set $\sigma_{\rm sys}=0\farcs2$ (per axis).
Column~7 gives the radial uncertainty $\epsilon_{99}=3.03\: \sigma_{x}$ in the X-ray position---i.e., $\chi^{2}_{2}=9.21=3.03^2$ corresponds to 99\% confidence on 2 degrees of freedom, for inclusion of the true source position.
Finally, columns 8--9 report estimates for the X-ray flux (\S\ref{s:analysis_flux}).

\subsection{Spectral Analysis\label{s:analysis_spec}}

\begin{figure}
\begin{center}
\epsfig{file=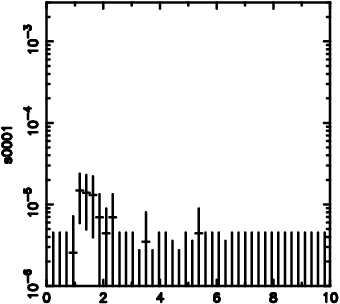,width=4.cm,angle=-90}
\epsfig{file=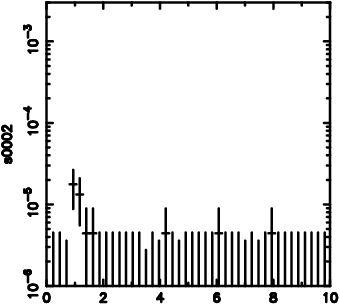,width=4.cm,angle=-90}
\epsfig{file=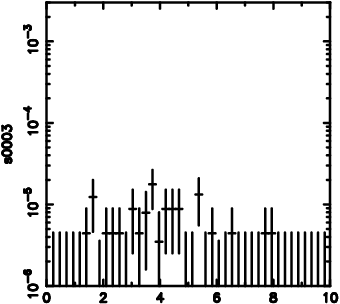,width=4.cm,angle=-90}
\epsfig{file=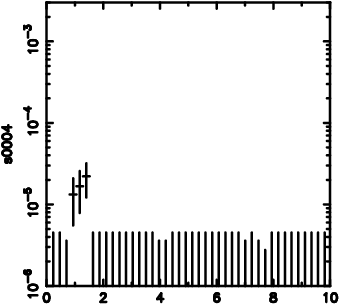,width=4.cm,angle=-90}
\epsfig{file=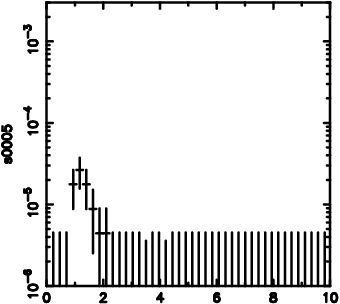,width=4.cm,angle=-90}
\epsfig{file=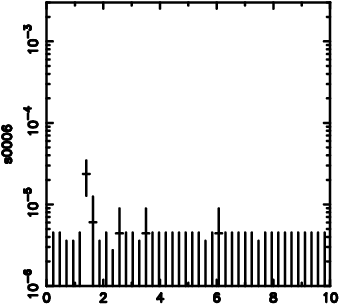,width=4.cm,angle=-90}
\epsfig{file=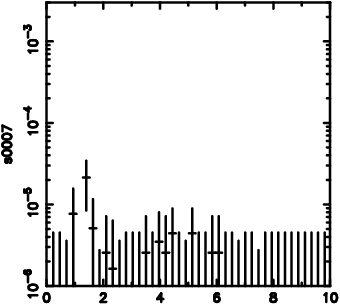,width=4.cm,angle=-90}
\epsfig{file=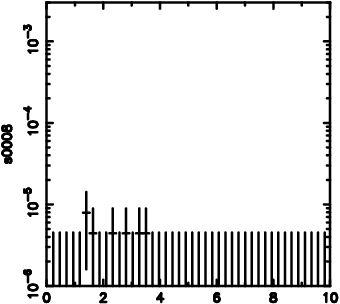,width=4.cm,angle=-90}
\epsfig{file=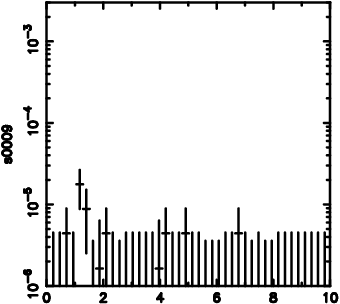,width=4.cm,angle=-90}
\epsfig{file=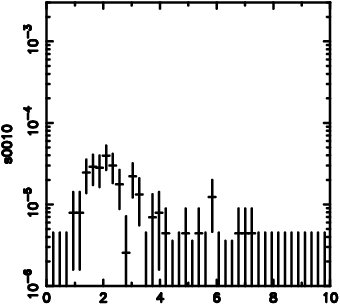,width=4.cm,angle=-90}
\epsfig{file=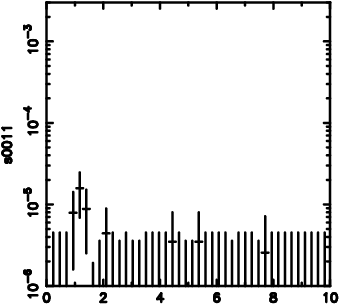,width=4.cm,angle=-90}
\epsfig{file=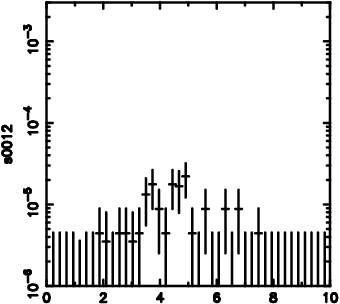,width=4.cm,angle=-90}
\epsfig{file=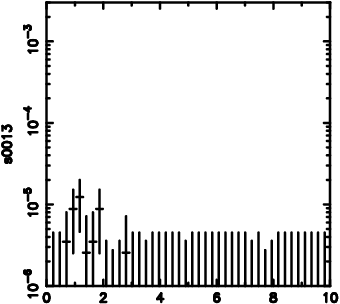,width=4.cm,angle=-90}
\epsfig{file=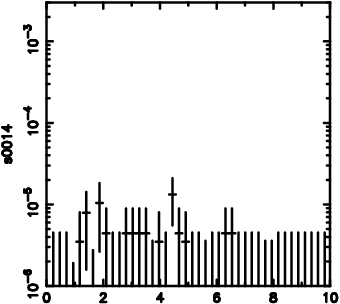,width=4.cm,angle=-90}
\epsfig{file=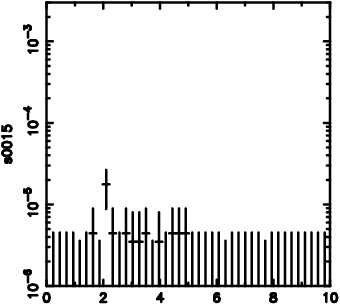,width=4.cm,angle=-90}
\end{center}
\figcaption{X-ray spectra for sources S01--S15, showing count rate per 0.234-keV bin versus energy (in keV).
Arranged left-to-right, S01 is at the upper left and S15, lower right.
\label{f:spectra_1}}
\end{figure}

Figures~\ref{f:spectra_1} and~\ref{f:spectra_2} display raw X-ray spectra of the 30 detected sources.
None of the sources has sufficient counts to warrant spectral analysis.
From the {\sc Hi} in the Galaxy \citep{dic90}, we compute a column density $N_{\rm H} \approx 1.4\times 10^{22}\, {\rm cm}^{-2}$ through the Galaxy in the $\gamma$-Cygni direction. 
Due to the significant drop in the \Chandra\ response above the mirror coating's iridium-M edges ($\approx\!2$~keV), any source with a substantial fraction of its detected photons above 2~keV is especially interesting.
Sources S03, S12, and S22 exhibit a spectrum with over a third of its counts above 2 keV, indicating an intrinsically hard or heavily absorbed X-ray source.
Consequently, each is a candidate (background) active galactic nucleus (AGN).
Such hard or heavily absorbed X-ray spectra seem to eliminate S03, S12, and S22 as the X-ray source associated with \EG---especially if the \gray\ source is a spin-powered pulsar.

\subsection{Flux estimates\label{s:analysis_flux}}

Column~8 of Table~\ref{t:data_x} presents our estimate of the X-ray photon spectral flux $K_{E}(E)$ at $E_{x} = 1\: {\rm keV}$ for each detected source.
Because the paucity of counts precludes a meaningful spectral analysis (\S\ref{s:analysis_spec}), we obtained these estimates using a fixed spectral shape---namely, a power law with photon index $\Gamma = 1.7$ through a column $N_{\rm H} = 1\!\times\!10^{22}\, {\rm cm}^{-2}$---for all sources.
The value of $N_{\rm H}$ is an estimate of the column density to the $\gamma$-Cyg SNR (\S\ref{s:oi_obs}).
While the typical column density through the Galaxy in this direction is about 1.5 times that value; absorption is likely to be quite patchy---especially, in the presence of the SNR-associated molecular cloud.
Freezing $\Gamma$ and $N_{\rm H}$, we used the XSPEC (v.11.3.2) spectral-fitting package \citep{arn96} to determine the normalization---$K_{E}$ at 1\,keV---and its statistical error for each source.
For computing the X-ray absorption, we utilized abundances (XSPEC's {\tt wilm}) from \citet{wil00} with cross-sections ({\tt vern}) from \citet{ver93} and allowed for interstellar extinction by grains using the model ({\tt tbabs}) of \citet{wil00}.
Column~9 calculates the (unabsorbed) energy flux $S$ in the 0.5--8-keV band.
To facilitate comparison with radiation in other bands, column~10 computes the (unabsorbed) energy spectral flux (flux density) $S_{\nu}(\nu)$ at $\nu_{x} = 242\: {\rm PHz}$ ($E_{x} = 1\,{\rm keV}$).

\subsection{Temporal Variability\label{s:analysis_var}}

\begin{figure}
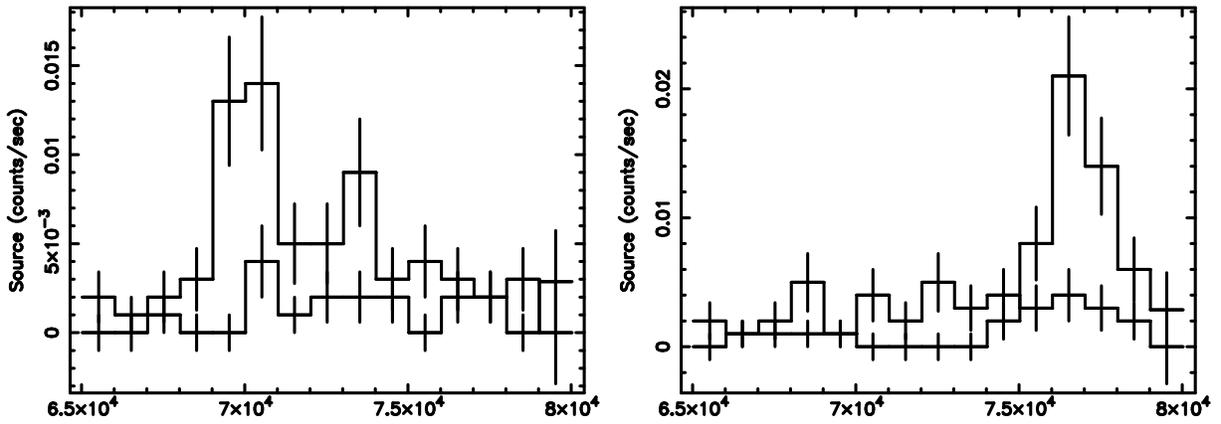

\begin{center}
\epsfig{file=f4a.eps,width=5.5cm,angle=-90}
\epsfig{file=f4b.eps,width=5.5cm,angle=-90}
\figcaption{Evidence for X-ray variability in two sources in the \EG\ field.
Plots show count rate versus time (in seconds) in the target (source extraction) region and in the reference (background extraction) region for sources S10 (left) and for S25 (right).
\label{f:flares}}
\end{center}
\end{figure}

The paucity of counts also precludes a sensitive variability analysis.
Nonetheless, the two X-ray-brightest sources---S10 and S25---exhibit evidence for temporal variations (Figure~\ref{f:flares}), suggestive of stellar coronal emission.
(Note that the $\alpha_{ix}$ is flatter than the other coronal-emission candidates.)
For S10, a likely 2MASS candidate counterpart (Table~\ref{t:data_oi}) reinforces this interpretation.
Such flares seem to eliminate S10 and S25 as the X-ray source associated with \EG---especially if the \gray\ source is a pulsar.

\section{Candidate Optical and Near-Infrared Counterparts\label{s:oi}}

\begin{figure}
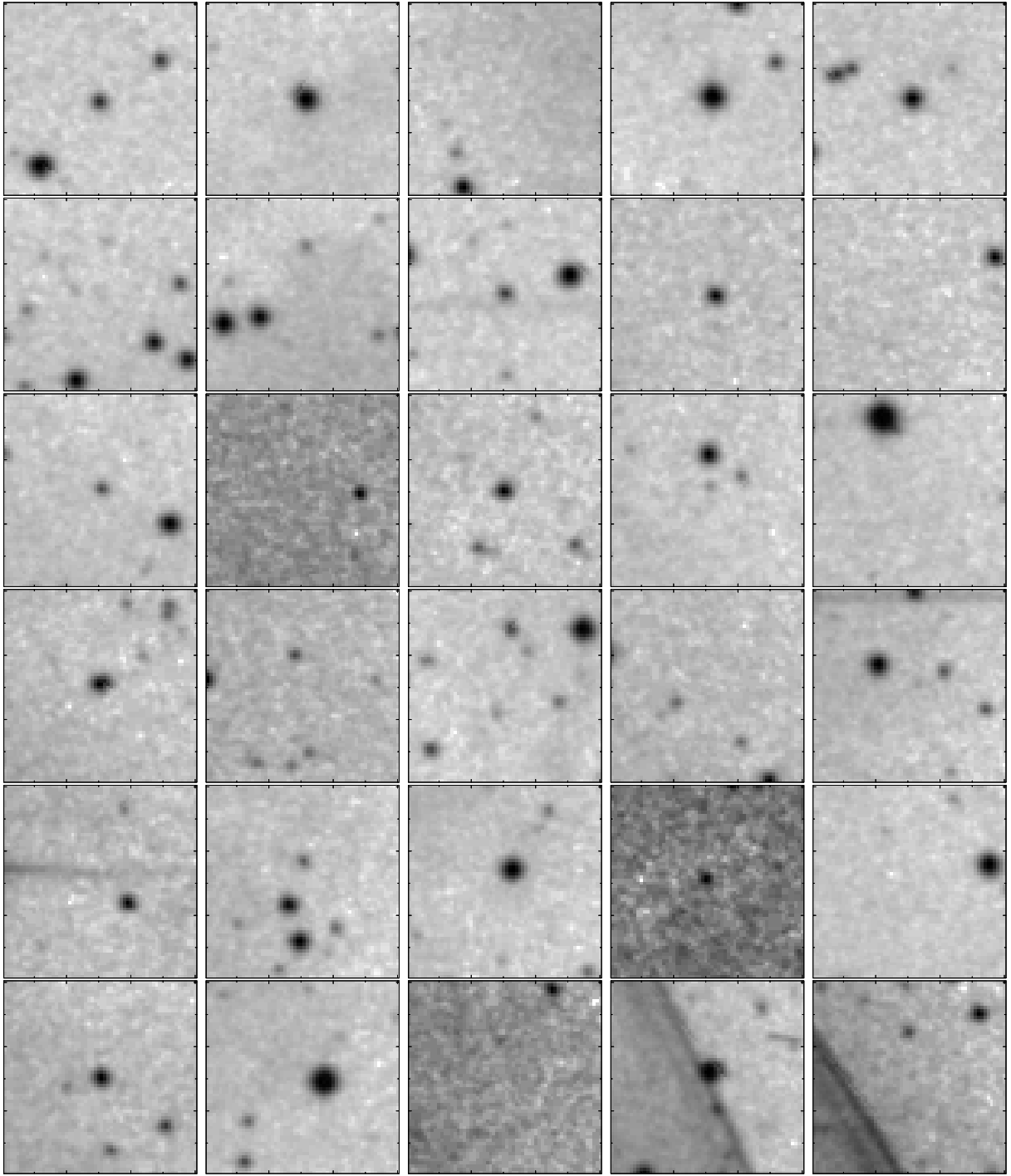

\begin{center}
\epsfig{file=f5a.ps,width=3.1cm,angle=-90}
\epsfig{file=f5b.ps,width=3.1cm,angle=-90}
\epsfig{file=f5c.ps,width=3.1cm,angle=-90}
\epsfig{file=f5d.ps,width=3.1cm,angle=-90}
\epsfig{file=f5e.ps,width=3.1cm,angle=-90}
\epsfig{file=f5f.ps,width=3.1cm,angle=-90}
\epsfig{file=f5g.ps,width=3.1cm,angle=-90}
\epsfig{file=f5h.ps,width=3.1cm,angle=-90}
\epsfig{file=f5i.ps,width=3.1cm,angle=-90}
\epsfig{file=f5j.ps,width=3.1cm,angle=-90}
\epsfig{file=f5k.ps,width=3.1cm,angle=-90}
\epsfig{file=f5l.ps,width=3.1cm,angle=-90}
\epsfig{file=f5m.ps,width=3.1cm,angle=-90}
\epsfig{file=f5n.ps,width=3.1cm,angle=-90}
\epsfig{file=f5o.ps,width=3.1cm,angle=-90}
\epsfig{file=f5p.ps,width=3.1cm,angle=-90}
\epsfig{file=f5q.ps,width=3.1cm,angle=-90}
\epsfig{file=f5r.ps,width=3.1cm,angle=-90}
\epsfig{file=f5s.ps,width=3.1cm,angle=-90}
\epsfig{file=f5t.ps,width=3.1cm,angle=-90}
\epsfig{file=f5u.ps,width=3.1cm,angle=-90}
\epsfig{file=f5v.ps,width=3.1cm,angle=-90}
\epsfig{file=f5w.ps,width=3.1cm,angle=-90}
\epsfig{file=f5x.ps,width=3.1cm,angle=-90}
\epsfig{file=f5y.ps,width=3.1cm,angle=-90}
\epsfig{file=f5z.ps,width=3.1cm,angle=-90}
\epsfig{file=f5aa.ps,width=3.1cm,angle=-90}
\epsfig{file=f5ab.ps,width=3.1cm,angle=-90}
\epsfig{file=f5ac.ps,width=3.1cm,angle=-90}
\epsfig{file=f5ad.ps,width=3.1cm,angle=-90}
\caption{Digitized Sky Survey (DSS) 2 red images centered at the position of each X-ray source listed in Table~\ref{t:data_x}.
Each image is $1\arcmin\!\times\!1\arcmin$, with North up and East left.
Arranged left-to-right, S01 is at the upper left and S30, lower right. 
For the last two fields, note the diffraction ring from the bright (V $\approx 2^{\rm mag}$) foreground star $\gamma$ Cygni (Sadir).\label{f:dss2}}
\end{center}
\end{figure}

Next we searched for candidate optical counterparts to the detected X-ray sources.
Due to the low X-ray flux of each \Chandra-detected source, a strong identification of an optical object with the X-ray source essentially excludes it as a candidate counterpart to \EG\ (\S\ref{s:summary}).
Figure~\ref{f:dss2} displays a Digitized Sky Survey (DSS) 2 red image centered on the coordinate of each of the 30 X-ray sources we detected.
Section~\ref{s:oi_cat} presents the results of our comparison with optical and near-infrared catalogs.
Section~\ref{s:oi_obs} discusses optical imaging and spectroscopic observations we performed to identify candidate optical counterparts to the X-ray sources.

\subsection{Comparison with Cataloged Point Sources\label{s:oi_cat}}

We used HEASARC's {\tt BROWSE}\footnote{See
http://heasarc.gsfc.nasa.gov/db-perl/W3Browse/w3browse.pl.} 
feature to search for cataloged objects within the 99\%-confidence radius ($\epsilon_{99}$) of X-ray source positions in Table~\ref{t:data_x}.
Table~\ref{t:data_oi} tabulates results of a cross correlation of the X-ray positions of the \Chandra-detected sources (column 1) 
with optical sources (columns~2--7) in the USNO-B1.0 catalog \citep[\S\ref{s:oi_usno};][]{mon03} and with near-infrared sources (columns~8--12) in the 2MASS catalog \citep[\S\ref{s:oi_2mass};][]{skr06}.

\subsubsection{USNO-B1.0\label{s:oi_usno}}

For fifteen (15) X-ray sources, we found a USNO-B1 (optical) source within the 99\%-confidence radius $\epsilon_{99}$ of the \Chandra\ position (Table~\ref{t:data_x}).
Of these, S11 had a second USNO-B1 source (not included in Table~\ref{t:data_oi}) 7 times farther than the first source---50 times more likely to be a chance coincidence based on the separation alone.
Table~\ref{t:data_oi} columns~2--4 list, respectively, the USNO-B1 right ascension RA, declination Dec, and RMS positional error $\sigma_{o}$ in the form ($\sigma_{o}({\rm RA})$, $\sigma_{o}({\rm Dec})$).
Column~5 gives the angular separation $\delta_{ox}$ between optical and X-ray positions; column~6, the I-band magnitude.
Column~7 estimates the probability $p_o(\delta_{ox}, {\rm I})$ for a chance coincidence within the observed separation of an object brighter than the I magnitude of the optical candidate.
We determined this probability from the I-magnitude distribution of the 1231 USNO sources within $8\arcmin$ of the X-ray pointing direction.
We designate a potential optical counterpart to an X-ray source as a ``strong candidate'' only if the sample impurity---i.e., probability of chance coincidence---$p_o(\delta_{ox}, {\rm I})<1\%$.
Twelve (12) sources---those marked with an asterisk in column~7 of Table~\ref{t:data_oi}---satisfy this criterion.

\subsubsection{2MASS\label{s:oi_2mass}}

For eighteen (18) X-ray sources, we found a 2MASS (near-infrared) source within the 99\%-confidence radius $\epsilon_{99}$ of the \Chandra\ position (Table~\ref{t:data_x}).
Table~\ref{t:data_oi} columns~8 and 9 list, respectively, the 2MASS right ascension RA and declination Dec, each with an RMS positional error $\sigma_{i}\approx 0\farcs080$.
Column~10 gives the angular separation $\delta_{ix}$ between near-infrared and X-ray positions; column~11, the K$_{s}$-band magnitude.
Column~12 estimates the probability $p_i(\delta_{ix}, {\rm K}_{s})$ for a chance coincidence within the observed separation of an object brighter than the K$_{s}$ magnitude of the infrared candidate.
We determined this probability from the $K_s$-magnitude distribution of the 2158 2MASS sources within $8\arcmin$ of the X-ray pointing direction.
We designate a potential near-infrared counterpart to an X-ray source as a ``strong candidate'' only if the sample impurity---i.e., probability of chance coincidence---$p_i(\delta_{ix}, {\rm K}_{s})<1\%$.
Seventeen (17) sources---those marked with an asterisk in column 12 of Table~\ref{t:data_oi}---satisfy this criterion.
Note that the 2MASS set of 17 strong candidate counterparts contains the USNO-B1 set of 12 strong candidates (\S\ref{s:oi_usno}).

\begin{figure}
\epsfig{file=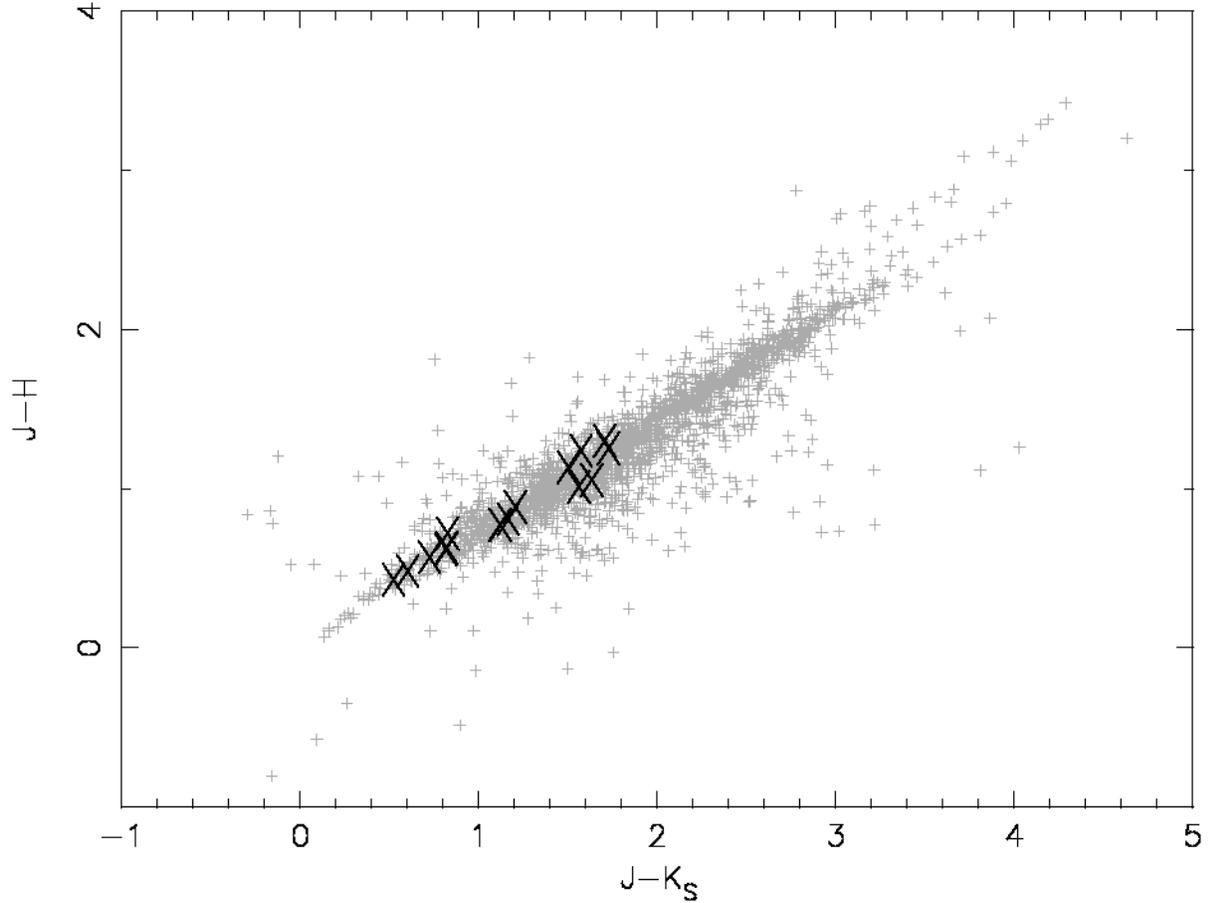,width=12cm,angle=-90}
\figcaption{Near-infrared color--color diagram for 2MASS objects within $8\arcmin$ from the center of the \EG\ field observed with \Chandra.
An ``X'' denotes a strong-candidate counterpart to an X-ray source (Table~\ref{t:2MASS_colors}).
Most field objects have colors consistent with reddened main-sequence stars, as do the objects identified with the \Chandra\ X-ray sources.
\label{f:2MASS_colors}}
\end{figure}

Table~\ref{t:2MASS_colors} tabulates the 2MASS near-infrared photometry (columns~2--7) of the 17 strong-candidate optical (visible--near-infrared) counterparts to \Chandra-detected X-ray sources.
Figure~\ref{f:2MASS_colors} is a near-infrared color--color diagram for all 2MASS sources within 8\arcmin\ of the pointing direction, denoting with an ``X'' those that are strong-candidate counterparts to the X-ray sources.
These candidate counterparts seem to be distributed as the field sources---i.e., as reddened main-sequence stars.
Although most Galactic-plane 2MASS objects are normal stars, the majority of objects identified with the X-ray sources need not be normal stars:
For example, the X-ray emission may originate in an accreting compact companion.

Column~8 of Table~\ref{t:2MASS_colors} converts the tabulated K$_{s}$ magnitude to an energy spectral flux $S_{\nu}(\nu)$ at $\nu_{i} = 139\, {\rm THz}$ ($\lambda_{i} = 2.16\,\mu{\rm m}$).
Column~9 then combines this near-infrared spectral flux $S_{\nu_{i}}$ with the X-ray spectral flux $S_{\nu_{x}}$ (Table~\ref{t:data_x} column 9) to obtain the effective infrared--X-ray energy spectral index $\alpha_{ix} \equiv - \ln(S_{\nu_{x}}/S_{\nu_{i}})/\ln(\nu_{x}/\nu_{i})$. 

\subsection{Optical and Near-Infrared Observations\label{s:oi_obs}}

In an effort to find additional candidate counterparts to the \Chandra-detected X-ray sources and to characterize the optical spectra of candidates, we performed optical and/or infrared observations. 
We obtained photometric imaging and dispersive spectroscopy at the Observatorio Astrof\'{\i}sico Guillermo Haro (\S\ref{sss:oagh}) and K$_{s}$ photometric imaging at Mount Palomar (\S\ref{sss:pal}).

\subsubsection{Observations at Observatorio Astrof\'{\i}sico Guillermo Haro\label{sss:oagh}}

Using the 2.12-m telescope of the Observatorio Astrof\'{\i}sico Guillermo Haro (in the Mexican side of the Sonora-Arizona region), we obtained optical imaging and spectroscopic observations at the locations of all X-ray sources in the field except S19.
In addition, we performed limited near-infrared (J and H) imaging of a few sources.
Table~\ref{t:obs_log} is a log of our optical observations, which employed the following instrumentation (described in more detail at www.inaoep.mx/$\sim$astrofi/cananea):

\begin{description}
\item The {\em Landessternwarte Faint-Object Spectrograph Camera} (LFOSC) provided initial optical imaging and spectroscopy, on 2005 May 30 and 31.
It is a user-friendly instrument that offers low-dispersion (5.6\,\AA/pixel) spectroscopy and quasi-simultaneous imaging over a 
$6\arcmin\!\times\!10\arcmin$ field with 0\farcs98/pixel plate scale.
\item The {\em Boller and Chivens spectrograph}\ gave optical spectra over the range 3500--6500\,\AA, with a typical dispersion of 3.5\,\AA/pixel, on 2005 June 2, 3, and 7.
\item The {\em C\'{a}mara Directa} $1024\!\times\!1024$ CCD camera provided VRI-filtered images over a $6\farcm7\!\times\!6\farcm7$ field with 0\farcs392/pixel plate scale, on 2005 June 4 and 5.
\item The {\em Cananea Near-Infrared Camera} (CANICA) furnished JH imaging deeper than 2MASS, for a limited number of sources, in late 2005 June and early 2006 June.
\end{description}

We reduced the acquired data with IRAF, using routines from NOAO's {\tt imred}, {\tt daophot}, {\tt twodspec} and {\tt onedspec} packages.
For astrometry, we matched {\tt daofind} physical positions with USNO-B celestial positions to within 0\farcs2 RMS, after adjusting plate scale and CCD rotation angle ($\approx\! 1.5\degr$).
Except for an unexpected warming of the CCD during V-band imaging on June 4 and a slight grating misplacement for two spectroscopic runs, the observations proceeded nominally.
The former anomaly affected the CCD background and efficiency, leading us to discard all V images from that date.
The latter resulted in missing H$\alpha$, but still gave useful spectra over the range 3000--6000\,\AA.

\begin{figure}
\begin{center}
\epsfig{file=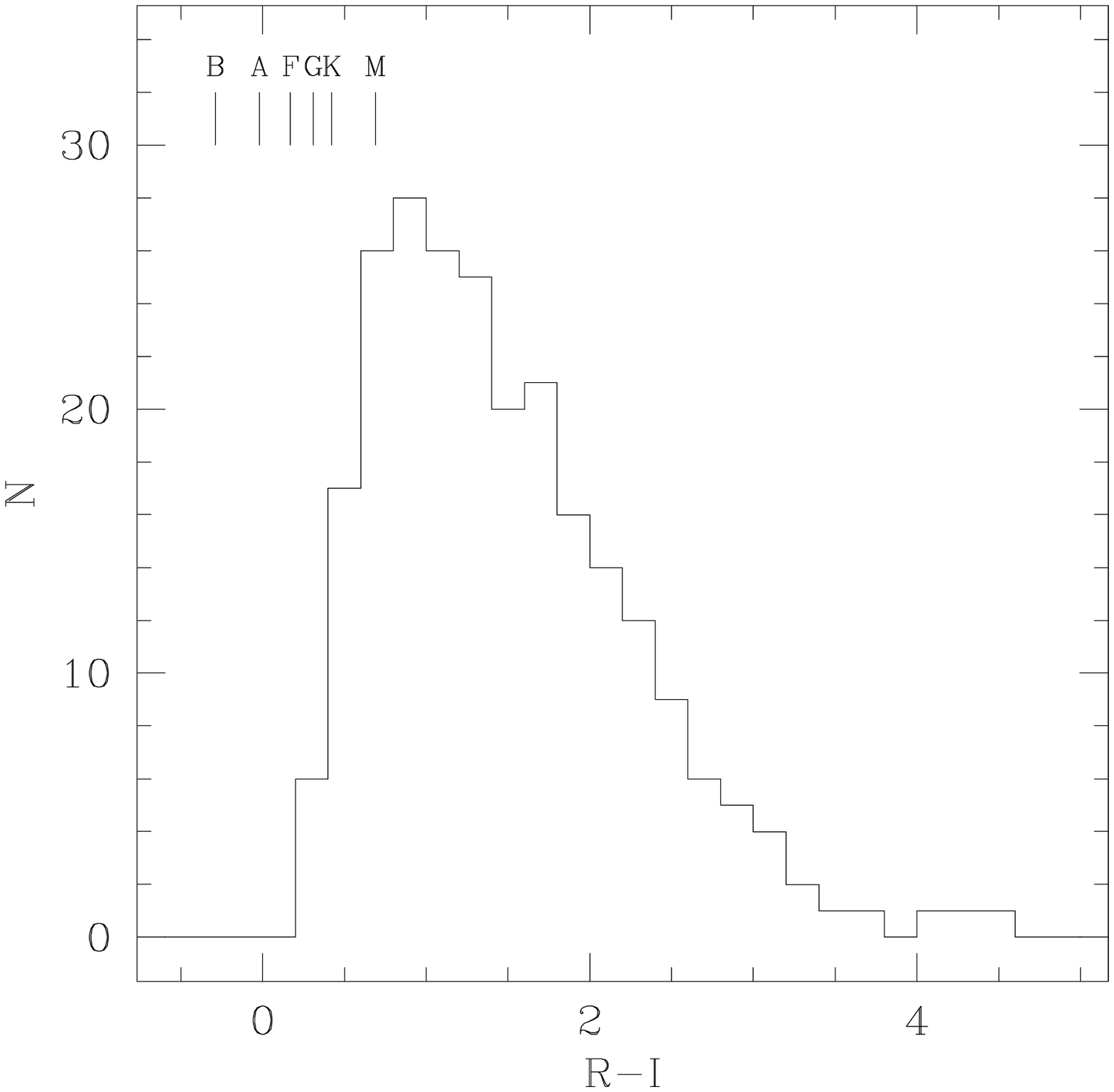, width=0.48\hsize}\hfill
\epsfig{file=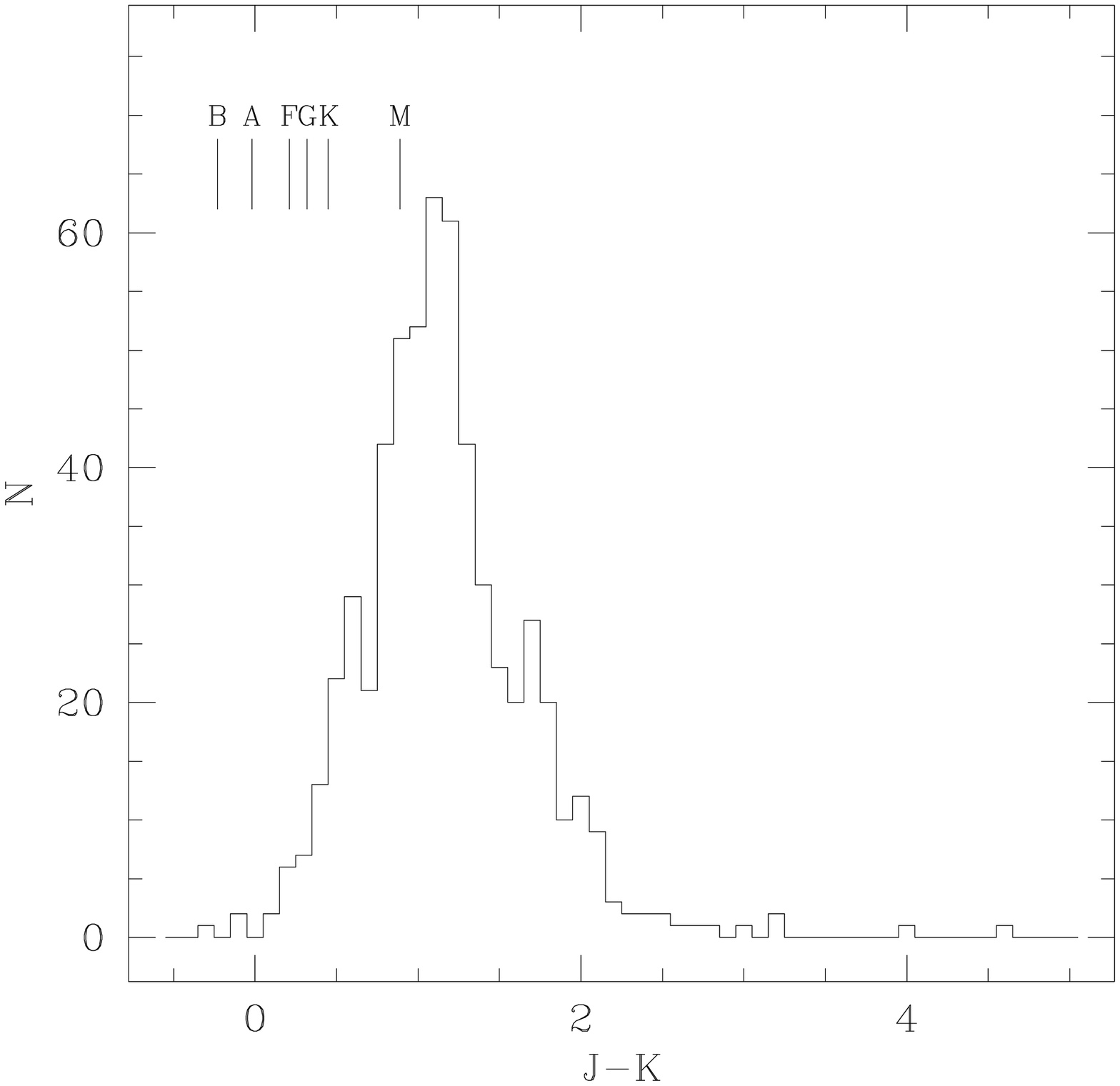, width=0.48\hsize}\hfill
\caption{Histograms of cataloged colors of field stars near \Chandra\ X-ray sources in the \EG\ field (Table~\ref{t:obs_log}).
Short vertical lines on top indicate unreddened colors for B0, A0, F0, G0, K0 and M0 main-sequence spectral types \citep{joh66}.
Left panel displays R-I colors; right panel, J-K colors.
\label{f:obs_colors}}
\end{center}
\end{figure}

Lacking photometry in a blue band, we cannot accurately measure extinction from our observations.
Using optical and radio data for the $\gamma$-Cyg SNR, \citet{joh74} estimated $A({\rm H}\alpha)\leq 3.7$, equivalent to $A({\rm V})\leq 4.6$, extrapolating \citet{mat90} extinction tables to ${\rm H}\alpha$ with $R_{\rm V}\equiv A({\rm V})/E({\rm B\!-\!V})=3.1$. 
Figure~\ref{f:obs_colors} shows the distribution of ${\rm R\!-\!I}$ and ${\rm J\!-\!K}$ colors for most stars in the observed fields.
With a mean $\langle {\rm R\!-\!I}\rangle =1.50$, the distribution is clearly redder than the intrinsic color of normal stars \citep{joh66}, even M stars ($\langle {\rm R\!-\!I}\rangle \approx 0.70$).
Assuming an average color excess $E({\rm R\!-\!I})\ga 1.2$, we find $A({\rm V})\ga 4.5$, roughly consistent with the previous bound \citep{joh74} and the hydrogen ({\sc Hi}+{\sc Hii}+2H$_{2}$) column $N_{\rm H}\ga 8.5\!\times\!10^{21}\,{\rm cm}^{-2}$.
Recently, \citet{mav03} presented imaging and long-slit spectroscopy of the $\gamma$-Cyg SNR, reporting  $N_{\rm H}= (6\pm1)\!\times\!10^{21}\,{\rm cm}^{-2}$---about two-thirds the typical total Galactic value $N_{\rm H}\approx 1.4\!\times\!10^{22}\, {\rm cm}^{-2}$ \citep{dic90} in this direction.
In supporting this estimate to the $\gamma$-Cyg SNR, \citet{mav03} cites CO maps of equivalent H$_{2}$ column $2\!\times\!10^{21} \,{\rm cm}^{-2}$ with a maximum $7\!\times\!10^{21}\,{\rm cm}^{-2}$ \citep{dam01} and an {\sc Hi} column $6\!\times\!10^{21}\,{\rm cm}^{-2}$ for the Galaxy from the maps of \citet{har97} with a maximum $8\!\times\!10^{21} \,{\rm cm}^{-2}$.
In \citet{bec04}, we used $N_{\rm H} = 1.4\!\times\!10^{22}\,{\rm cm}^{-2}$, implying $A_{{\rm V}}= 7.4$ and $E({\rm R\!-\!I})\ga 2.0$---more representative of a typical column through the Galaxy in that direction.
Accordingly, we set the column at the $\gamma$-Cyg SNR to $N_{\rm H} = 1.0\!\times\!10^{22}\,{\rm cm}^{-2}$ for the current analysis  (\S\ref{s:analysis_flux}).


Table~\ref{t:obs_data} tabulates the results of our optical imaging and spectroscopy of \Chandra-detected sources (column 1).
Column 2 identifies each candidate counterpart as a USNO(B1) or 2MASS object, parenthesizing those that are not strong candidates---i.e., those with the probability of chance coincidence $p(\delta)\!\ga\!1\%$ (\S\ref{s:oi_cat}).
For sources without cataloged candidate counterparts, column 2 states whether the target was ``Unobserved'' or observed with no candidate (``None'') or with an uncataloged candidate (``New'') identified.
Columns 3--6 list results of photometry; column 7, those of spectroscopy.
Column 8 provides comments---including spectral-type of identified stars, parenthesized probability of chance coincidence for marginal (not ``strong'') candidates, and offset from the X-ray position of any uncataloged object we identified.

Our imaging observations discovered uncataloged (``New'') candidate counterparts to \Chandra\ X-ray sources S23 (closer than the USNO, 2MASS object) and S25.
However, due to the moderately large probability of chance coincidence, we cannot regard either of these as ``strong candidates''.
In addition, these observations established limits on optical or near-infrared magnitudes of undetected counterparts within $3\arcsec$ of X-ray sources S03, S12, S15, S17, S18, S20, S22, S27 and S30.
Our limits are more sensitive than the USNO-B1 completeness limit by $\approx\! 2^{\rm mag}$; than the 2MASS, by $\approx\! 3^{\rm mag}$.

\subsubsection{Observations at Palomar Observatory\label{sss:pal}}

We obtained $K_s$-band images of a portion of the \Chandra\ field with the Wide-field Infrared Camera \citep[WIRC;][]{wil03} on the 200-inch (5-m) telescope of the Palomar Observatory, operated by the California Institute of Technology.
We observed a number of pointings with 9-minute WIRC exposures.
For the data reduction, we employed custom PyRAF routines, subtracting dark frames, then producing a sky frame for subtraction by taking a sliding box-car window of four exposures on either side of a reference exposure.
After adding together these exposures, we identified all the stars and produced masks for the stars that were used to improve the sky frames in a second round of sky subtraction.
Astrometric and photometric referencing used $\approx\! 500$ unsaturated 2MASS stars at each pointing, where astrometric uncertainties are dominated by the $\approx\! 0\farcs08$ (per axis) absolute uncertainty of the 2MASS data and photometric uncertainties are $\approx\! 0.03$~mag.

\begin{figure}
\begin{center}
\epsfig{file=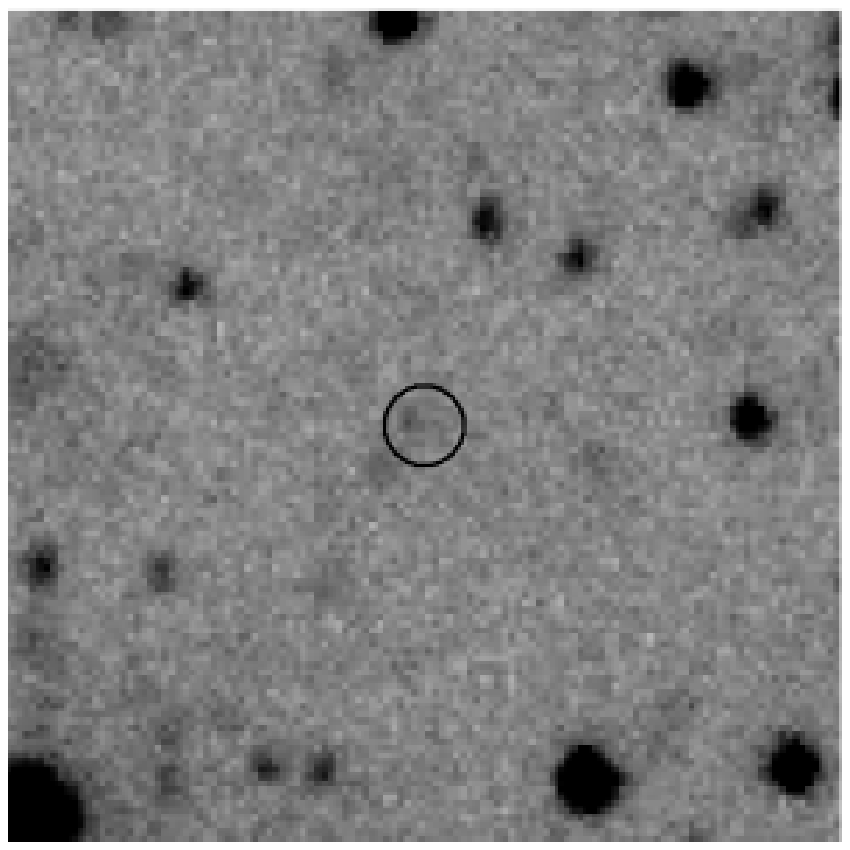,width=5.4cm,angle=0}
\epsfig{file=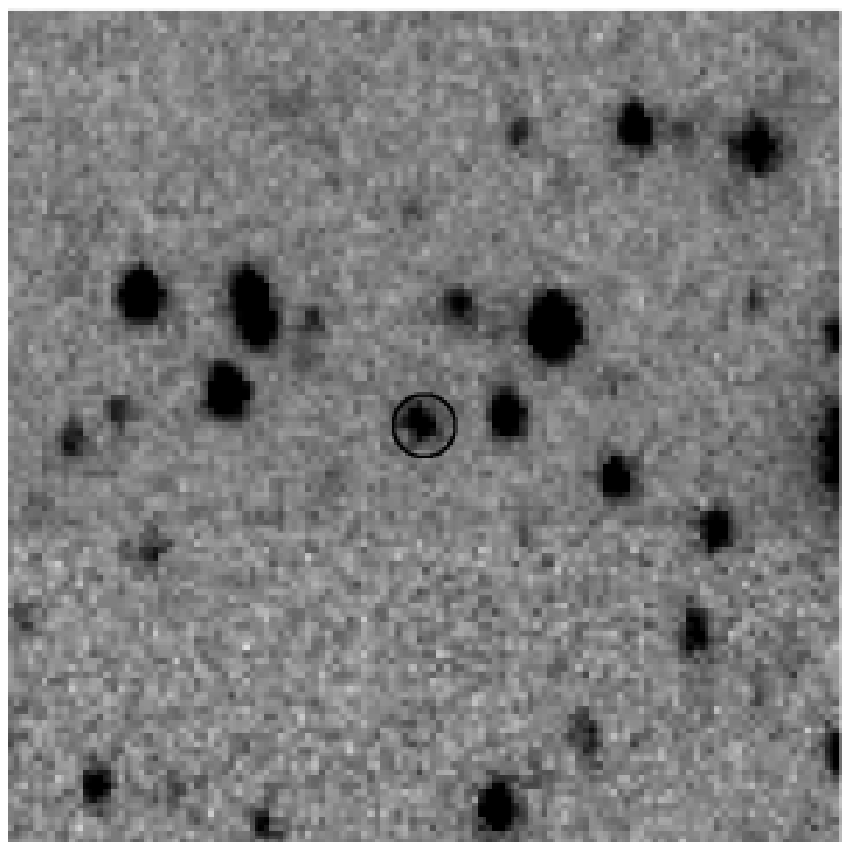,width=5.4cm,angle=0}
\epsfig{file=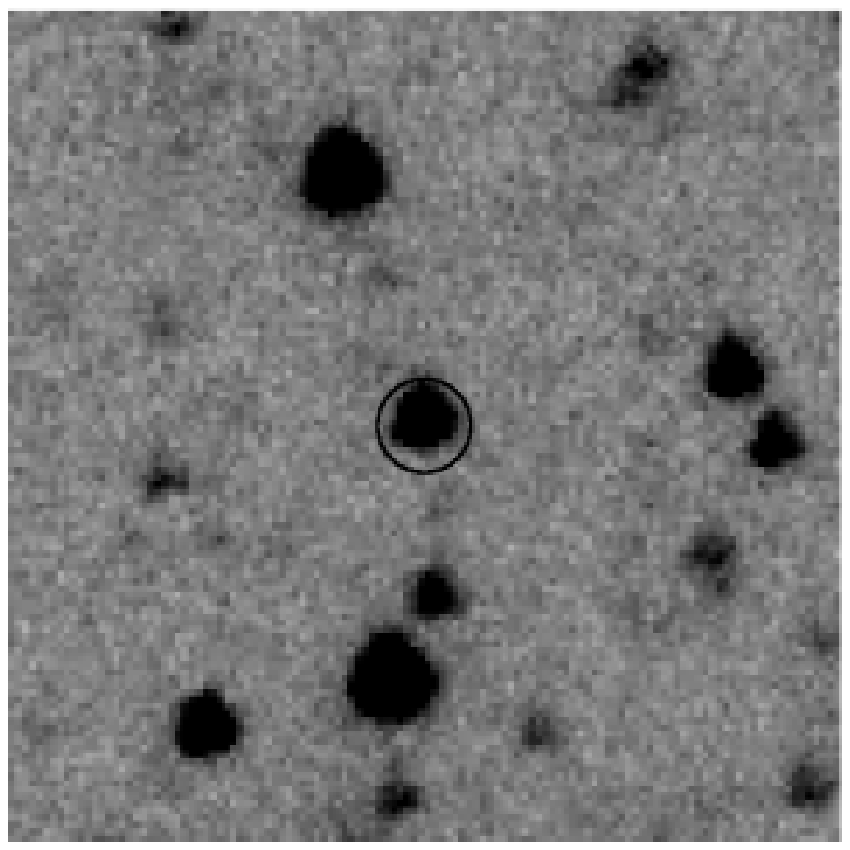,width=5.4cm,angle=0}
\epsfig{file=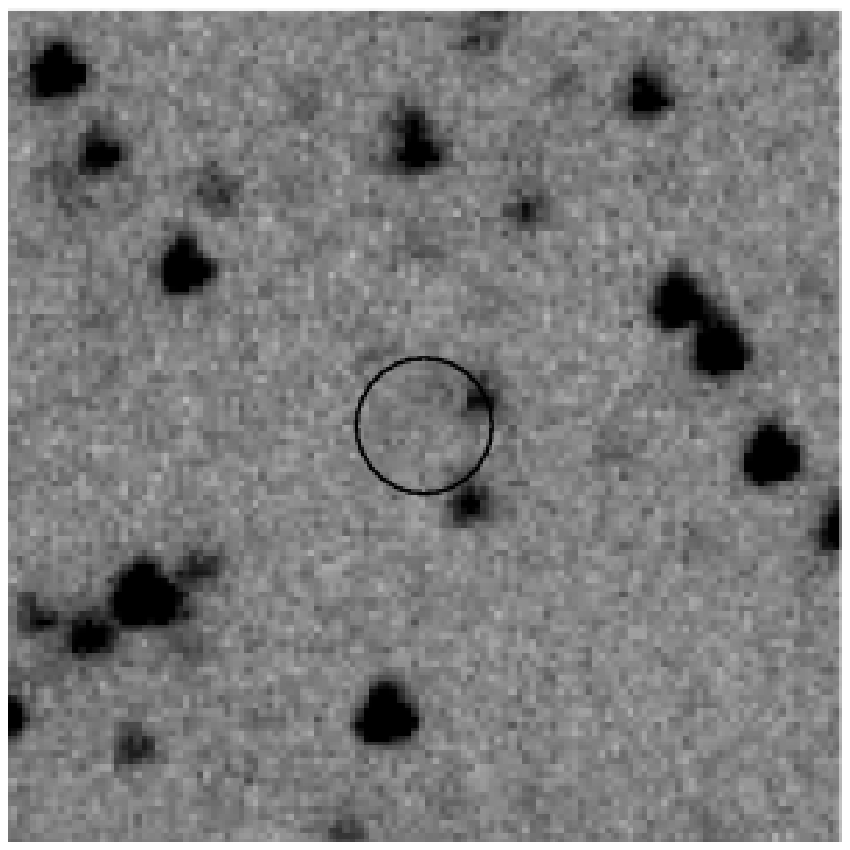,width=5.4cm,angle=0}
\epsfig{file=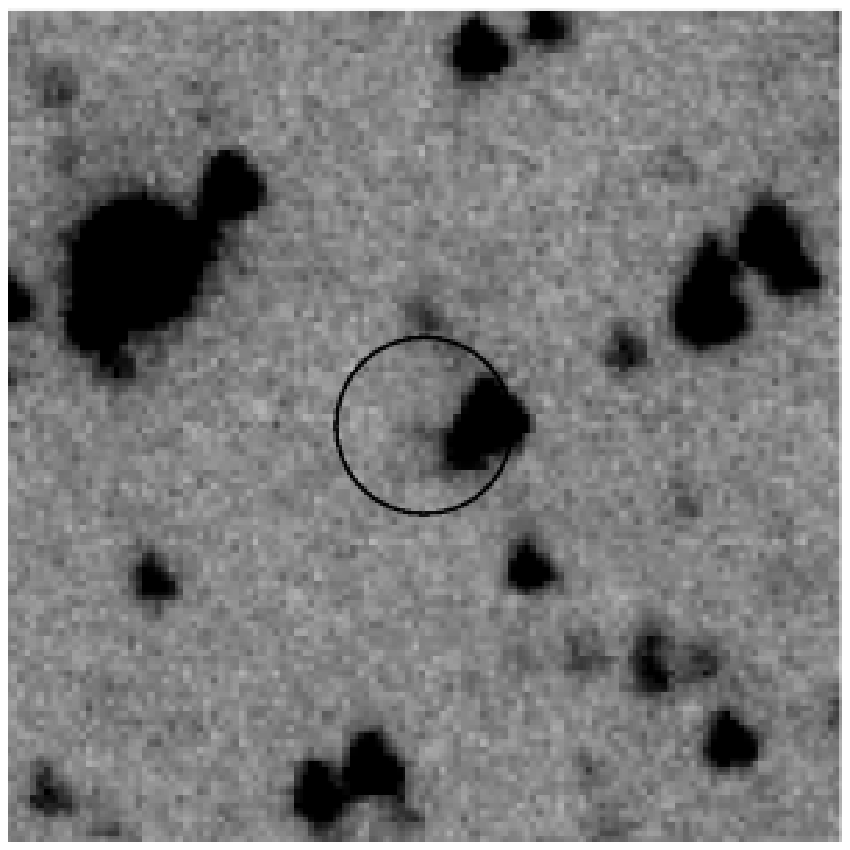,width=5.4cm,angle=0}
\epsfig{file=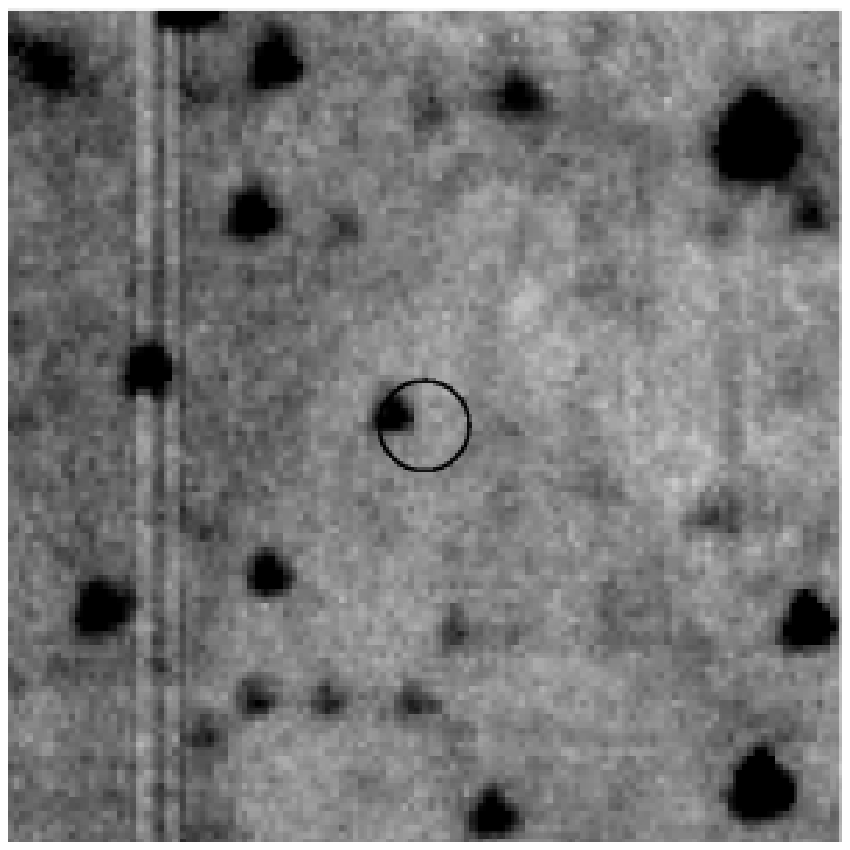,width=5.4cm,angle=0}
\epsfig{file=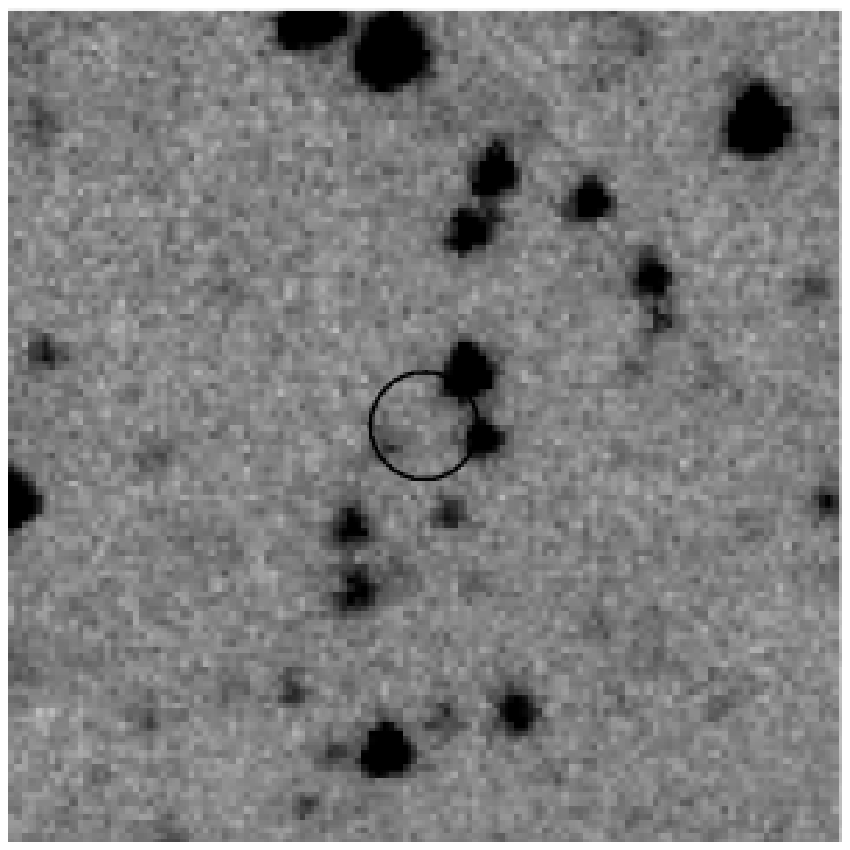,width=5.4cm,angle=0}
\caption{WIRC K$_s$ images of fields listed in Table~\ref{t:wirc_data}, each centered on the position (Table~\ref{t:data_x}) of the X-ray source---S03, S12, S17, S19, S20, S25, and S30.
Each image is $30\arcsec\!\times\!30\arcsec$, with North up and East left.
Superposed on each image is the 99\%-confidence error circle for the \Chandra\ source position.
Arranged left-to-right, S03 is at the upper left and S30, lower right.
S12, S17, and S20 have strong-candidate K$_s$ counterparts (see Table~\ref{t:wirc_data} column 6).\label{f:wirc}}
\end{center}
\end{figure}

These observations were $\approx\! 5^{\rm mag}$ deeper than the 2MASS limit and covered the positions of all the detected X-ray sources (Table~\ref{t:data_x}) except S01, S06, S07, S11, S13, S15, S18, S22, S23, and S27.
In addition to the candidate counterparts listed in Table~\ref{t:data_oi}, we found (Table~\ref{t:wirc_data}) seven (7) more candidates counterparts---namely for, S03, S12, S17, S19, S20, S25, and S30---and established a deeper faintness limit for S21 (K$_{s} > 19.7$).
Figure~\ref{f:wirc} displays $K_{s}$ images of the fields containing these sources.
Three (3) of the potential candidates---S12, S17, and S20---are ``strong''.
Consequently, combining our WIRC results (Table~\ref{t:wirc_data}) with the 2MASS comparison (Table~\ref{t:data_oi}), 20 of the 30 \Chandra-detected x-ray sources have strong-candidate K-band counterparts.

All the seven (7) uncataloged candidates we found using the WIRC at Palomar observatory were fainter than the 2MASS K-band completeness limit.
Five (5) of these were much (about 3--5 magnitudes) fainter, and thus understandably undetected in the observations at the Observatorio Astrof\'{\i}sico Guillermo Haro (\S\ref{sss:oagh} and Table~\ref{t:obs_data}).
In fact those observations did not search one of the locations (S19).
For the second brightest (K = 16.9) WIRC potential counterpart (S25, {\em not} a strong candidate), the Guillermo-Haro observations did not detect the object in R or I, but did in H.
For the brightest (K = 15.3) potential counterpart (S17, a strong candidate), they did not detect the object in R or I, and there were no near-infrared measurements.

\section{Discussion and Summary\label{s:summary}}

Using the \Chandra\ X-ray Observatory, we continued our search \citep{bec04} for possible X-ray counterparts to the intriguing \gray\ source \EG\ (\CG).
We found 30 X-ray sources in a field centered on the 3EG-cataloged position, located within the $\gamma$-Cyg SNR.
One of these---S10---lies $2\farcs3\!\pm\!1\farcs0$ from a previously detected source---S211 of \citet{bec04}---and could be the same source.
Most \Chandra\ sources in the field are naturally near the completeness limit  ($\approx\!10$ counts), slightly above the detection limit ($\approx\!7$ counts).
The strongest source---S10---is about 10 times the detection limit.
For the 14.3-ks observing time, detected sources have (\S\ref{s:analysis_flux} and Table~\ref{t:data_x}) photon spectral fluxes $K_{E} \ga 1.6\!\times\!10^{-6}$~photons/(cm$^2$ s keV) at 1~keV, X-ray energy fluxes $S \ga 9\!\times\!10^{-15}$~erg/(cm$^2$ s) in the 0.5--8-keV band, and energy spectral fluxes (flux densities) $S_{\nu} \ga 1.1$~nJy at 242~PHz (1~keV).
For sources at distances comparable to that of the $\gamma$-Cyg SNR (about 1.6~kpc), the X-ray luminosity would be $L_{x} > 3\!\times\!10^{30}$~erg/s in the 0.5--8-keV band.

\citet{zav04} show that the effective optical--X-ray energy index $\alpha_{ox} \approx 0.35\pm 0.1$---corresponding to $\log(L_{x}/L_{o}) \approx +2$---for spin-powered pulsars detected in both these bands.
Hence, if one of the X-ray sources in the field is a (similar) spin-powered pulsar, its flux density in an optical (or near-IR) band would  be no brighter than about $0.1\, \mu$Jy ($\approx\!25^{\rm mag}$)---i.e., undetected in either the 2MASS or the USNO-B catalog or by our visible or near-infrared observations.
Consequently, those 20 X-ray sources with strong-candidate counterparts (Tables~\ref{t:data_oi} and \ref{t:wirc_data}) are {\em not} good candidates for spin-powered pulsars.
Our optical spectroscopy (\S\ref{s:oi_obs} and Table~\ref{t:obs_data}) shows that most strong-candidate counterparts are reddened main-sequence stars, with $\alpha_{ix} \approx 2$---corresponding to $\log(L_{x}/L_{i}) \approx -3$.
Furthermore, our X-ray observations of variability (\S\ref{s:analysis_var}) in S10 and S25 and of apparently highly-absorbed spectra (\S\ref{s:analysis_spec}) in S03, S12, and S22 would suggest that none of these sources is a spin-powered pulsar.

The available data do not disqualify the remaining 10 X-ray sources--- S07, S09, S15, S18, S19, S21, S23, S27, S29, and S30---as possible candidates for a spin-powered pulsar.
However, there is nothing distinguishing any of these X-ray sources from the others:
They could represent several classes of objects.
In fact, based upon deep surveys, we would expect approximately 10 extragalactic X-ray sources in this field.
As we (Becker \et\ 2004) reported previously, there is {\em no} radio evidence for a pulsar in the $\gamma$-Cyg SNR field, down to a limiting sensitivity of $L_{\nu} = 0.09$ mJy kpc$^{2}$ at $\nu_{r}=$ 820~MHz (for an assumed 4\% pulse duty cycle).
Thus, if one of the X-ray sources is a spin-powered pulsar, it would be rather radio-quiet, as is Geminga.
Finally, in that accurate determination of a \gray\ source position against a strong gradient in the diffuse background is difficult, we must also admit the possibility that the EG3 position for \EG\ is incorrect.
Finally we note that without high-precision X-ray spectra of each of the candidate X-ray sources, and detailed follow up in other wavelength bands, there is essentially no satisfactory way in which to eliminate some of the candidates from consideration. 
This is especially true for neutron stars as the the corresponding infrared-visible fluxes are very weak making such observatiobs especially difficult.
In such cases, the principal and important {\sl Chandra} contribution is to provide target lists as we have done with accurate positions as a basis for future studies.

\acknowledgments
Those of us (MCW, DAS, RFE, SLO, \& AFT) at NASA's Marshall Space Flight Center (MSFC) acknowledge support from the \Chandra\ Program. 
DLK acknowledges postdoctoral support through a Pappalardo Fellowship in Physics at MIT.
Our research utilized data products from the Two-Micron All Sky Survey (2MASS)---a joint project of the University of Massachusetts and the Infrared Processing and Analysis Center at the California Institute of Technology, funded by the NASA and by the NSF.
The research also used data from the USNOFS Image and Catalog Archive---operated by the United States Naval Observatory (USNO), Flagstaff Station (http://www.nofs.navy.mil/data/fchpix/).
In addition, we acknowledge use of NASA's Astrophysics Data System Bibliographic Service; of the VizieR Service at the Centre de Donn\'{e}es astronomiques de Strasbourg; and of the High-Energy Astrophysics Science Archive Research Center (HEASARC), provided by NASA's Goddard Space Flight Center (GSFC).

\begin{deluxetable}{crrrrrrccc}
\tabletypesize{\scriptsize}
\tablewidth{0pc}
\tablecaption{\Chandra\ X-ray sources in the \EG\ field. \label{t:data_x}}
\tablehead{(1) & \multicolumn{1}{c}{(2)} & \multicolumn{1}{c}{(3)} & \multicolumn{1}{c}{(4)} & \multicolumn{1}{c}{(5)} & \multicolumn{1}{c}{(6)} & \multicolumn{1}{c}{(7)} & (8) & (9) & (10)}
\startdata
 Source & \multicolumn{1}{c}{RA(J2000)} & \multicolumn{1}{c}{Dec(J2000)}
 & \multicolumn{1}{c}{$\theta_{\rm ext}$} & \multicolumn{1}{c}{$C_{x}$} 
 & \multicolumn{1}{c}{$\sigma_{x}$} & \multicolumn{1}{c}{$\epsilon_{99}$}
 & $K_{E}$\,@\,1\,keV & $S(0.5\!-\!8\, {\rm keV})$ & $S_{\nu}$\,@\,242\,PHz \\
 & \multicolumn{1}{c}{$^{\rm h}\: \ ^{\rm m}\ \ ^{\rm s}\: $\ \ \ \ }  
 & \multicolumn{1}{c}{\ $\ \arcdeg\: \ \ \arcmin\: \ \ \arcsec$\ \ \ }
 & \multicolumn{1}{c}{$\arcsec$} & & \multicolumn{1}{c}{$\arcsec$}  & \multicolumn{1}{c}{$\arcsec$} & $10^{-6} {\rm /(cm^{2}\, s\, keV)}$ & $10^{-15} {\rm erg/(cm^{2}\, s)}$ & nJy \\ \hline\\[-2ex]
S01 & 20 20 31.943 & +40 11 52.31 &  9.9 & 15 & 1.06 & 3.21 &  $4.3\pm 1.2$ & $24\pm 7$ &  $2.8\pm 0.8$ \\
S02 & 20 20 33.019 & +40 20 45.52 &  6.3 & 11 & 0.82 & 2.48 &  $2.4\pm 0.8$ & $13\pm 4$ &  $1.6\pm 0.5$ \\
S03 & 20 20 33.691 & +40 18 27.95 &  5.0 & 31 & 0.47 & 1.42 &  $6.5\pm 1.2$ & $37\pm 7$ &  $4.3\pm 0.8$ \\
S04 & 20 20 35.397 & +40 22 09.75 &  7.0 & 11 & 0.91 & 2.76 &  $2.4\pm 0.8$ & $14\pm 4$ &  $1.6\pm 0.5$ \\
S05 & 20 20 45.900 & +40 20 31.53 &  3.3 & 18 & 0.43 & 1.30 &  $4.7\pm 1.1$ & $27\pm 6$ &  $3.1\pm 0.7$ \\
S06 & 20 20 47.619 & +40 10 41.05 &  8.6 &  9 & 1.20 & 3.64 &  $2.0\pm 0.8$ & $11\pm 5$ &  $1.3\pm 0.5$ \\
S07 & 20 20 49.478 & +40 08 58.00 & 11.8 & 10 & 1.50 & 4.55 &  $2.2\pm 0.9$ & $13\pm 5$ &  $1.5\pm 0.6$ \\
S08 & 20 20 51.255 & +40 15 19.86 &  2.4 &  7 & 0.48 & 1.45 &  $1.6\pm 0.6$ &  $9\pm 4$ &  $1.1\pm 0.4$ \\
S09 & 20 20 52.808 & +40 24 31.25 &  7.4 &  9 & 1.04 & 3.15 &  $5.9\pm 1.8$ & $33\pm10$ &  $3.9\pm 1.2$ \\
S10 & 20 20 54.884 & +40 24 19.02 &  6.9 & 63 & 0.46 & 1.39 & $16.7\pm 2.2$ & $94\pm12$ & $11.1\pm 1.4$ \\
S11 & 20 20 57.901 & +40 10 50.12 &  7.4 & 10 & 0.99 & 3.00 &  $2.2\pm 0.8$ & $12\pm 4$ &  $1.5\pm 0.5$ \\
S12 & 20 21 04.118 & +40 21 39.50 &  3.1 & 37 & 0.36 & 1.09 &  $9.8\pm 1.6$ & $55\pm 9$ &  $6.5\pm 1.1$ \\
S13 & 20 21 11.737 & +40 10 41.23 &  7.9 &  8 & 1.16 & 3.51 &  $2.4\pm 0.8$ & $13\pm 5$ &  $1.6\pm 0.5$ \\
S14 & 20 21 12.922 & +40 24 05.33 &  6.7 & 16 & 0.73 & 2.21 &  $4.2\pm 1.1$ & $24\pm 6$ &  $2.8\pm 0.7$ \\
S15 & 20 21 14.734 & +40 12 17.73 &  5.6 & 12 & 0.72 & 2.18 &  $2.8\pm 0.8$ & $16\pm 5$ &  $1.9\pm 0.6$ \\
S16 & 20 21 20.357 & +40 17 28.00 &  2.6 & 29 & 0.36 & 1.09 &  $8.3\pm 1.6$ & $47\pm 9$ &  $5.5\pm 1.0$ \\
S17 & 20 21 23.775 & +40 19 03.14 &  3.4 &  8 & 0.57 & 1.73 &  $1.9\pm 0.7$ & $11\pm 4$ &  $1.3\pm 0.5$ \\
S18 & 20 21 24.153 & +40 09 43.66 & 11.6 & 10 & 1.52 & 4.61 &  $1.8\pm 0.7$ & $10\pm 5$ &  $1.2\pm 0.6$ \\
S19 & 20 21 26.559 & +40 20 08.01 &  4.5 &  6 & 0.79 & 2.39 &  $1.6\pm 0.7$ &  $9\pm 4$ &  $1.1\pm 0.5$ \\
S20 & 20 21 26.684 & +40 14 52.56 &  4.9 & 10 & 0.69 & 3.21 &  $2.4\pm 0.7$ & $14\pm 5$ &  $1.6\pm 0.6$ \\
S21 & 20 21 30.655 & +40 26 46.39 & 15.2 & 36 & 1.06 & 2.79 & $10.2\pm 1.9$ & $58\pm11$ &  $6.8\pm 1.2$ \\
S22 & 20 21 30.785 & +40 12 17.97 &  8.7 & 16 & 0.92 & 3.97 &  $4.4\pm 1.1$ & $25\pm 6$ &  $2.9\pm 0.7$ \\
S23 & 20 21 31.010 & +40 11 19.63 & 10.2 & 10 & 1.31 & 3.67 &  $3.6\pm 1.1$ & $20\pm 6$ &  $2.4\pm 0.7$ \\
S24 & 20 21 31.344 & +40 22 56.66 &  8.4 &  8 & 1.21 & 2.24 &  $2.3\pm 0.8$ & $13\pm 4$ &  $1.5\pm 0.5$ \\
S25 & 20 21 32.442 & +40 24 59.18 & 12.0 & 50 & 0.74 & 1.58 & $14.6\pm 2.2$ & $82\pm12$ &  $9.7\pm 1.5$ \\
S26 & 20 21 33.154 & +40 15 57.65 &  5.9 & 31 & 0.52 & 4.55 &  $7.3\pm 1.4$ & $41\pm 8$ &  $4.8\pm 0.9$ \\
S27 & 20 21 33.586 & +40 10 35.93 & 12.2 & 11 & 1.50 & 3.88 &  $3.3\pm 1.2$ & $18\pm 7$ &  $2.2\pm 0.8$ \\
S28 & 20 21 35.060 & +40 23.58.81 & 11.0 & 12 & 1.28 & 3.70 &  $5.7\pm 1.5$ & $32\pm 8$ &  $3.8\pm 1.0$ \\
S29 & 20 21 46.353 & +40 17 58.27 & 10.1 & 12 & 1.22 & 3.85 &  $2.6\pm 0.9$ & $15\pm 5$ &  $1.8\pm 0.6$ \\
S30 & 20 21 47.128 & +40 18 50.10 & 10.6 & 12 & 1.27 & 1.91 &  $3.0\pm 1.1$ & $17\pm 6$ &  $2.0\pm 0.7$ \\
\enddata
\end{deluxetable}

\clearpage

\begin{deluxetable}{cccccccccccc}
\tabletypesize{\scriptsize}
\rotate
\tablewidth{0pc}
\tablecaption{Candidate cataloged counterparts to X-ray sources in the \EG\ field.\label{t:data_oi}}
\tablehead{(1) & (2) & (3) & (4) & (5) & (6) & (7) & (8) & (9) & (10) & (11) & (12)}
\startdata
  { } & \multicolumn{6}{c}{USNO (optical) candidate counterpart} 
  & \multicolumn{5}{c} {2MASS (infrared) candidate counterpart} \\ 
  Source & RA(J2000) & Dec(J2000) & $\sigma_{o}^a$ & $\delta_{ox}$ & I
  & $p_{o}(\delta_{ox},{\rm I})$
  & RA(J2000)$^b$ & Dec(J2000)$^b$ & $\delta_{ix}$ & K$_{s}$ & $p_{i}(\delta_{ix},{\rm K}_{s})$ \\
  {} & \multicolumn{1}{c}{$^{\rm h}\: \ ^{\rm m}\ \ ^{\rm s}\: $\ \ \ \ } 
  & \multicolumn{1}{c}{\ $\ ^\circ\: \ \ \arcmin\: \ \ \arcsec$\ \ \ }
  & $\arcsec$ & $\arcsec$  & mag & \%
  & \multicolumn{1}{c}{$^{\rm h}\: \ ^{\rm m}\: \ \ ^{\rm s}\: $\ \ \ \ } 
  & \multicolumn{1}{c}{\ $\ \arcdeg\: \ \ \arcmin\ \ \arcsec$\ \ \ } 
  & $\arcsec$ & mag & \% \\ \hline\\[-2ex]
 S01  & 20 20 32.013 & +40 11 51.10 & (0.174, 0.051) & 1.46 & 15.76 & 0.39*  & 20 20 32.019 & +40 11 50.86 & 1.71 & 13.33 & 0.70*  \\
 S02  & 20 20 32.962 & +40 20 45.51 & (0.433, 0.131) & 0.65 & 13.85 & 0.055* & 20 20 32.992 & +40 20 45.51 & 0.31 & 11.88 & 0.007* \\
 S04  & 20 20 35.358 & +40 22 10.66 & (0.051, 0.126) & 1.04 & 13.48 & 0.13*  & 20 20 35.340 & +40 22 10.25 & 0.83 & 11.82 & 0.05*  \\
 S05  & 20 20 45.905 & +40 20 31.48 & (0.063, 0.081) & 0.07 & 14.27 & 0.0007*& 20 20 45.888 & +40 20 31.17 & 0.38 & 12.05 & 0.012* \\
 S06  &              &              &                &      &       &        & 20 20 45.888 & +40 10 40.31 & 0.86 & 14.65 & 0.48*  \\
 S07  &              &              &                &      &       &        & 20 20 49.366 & +40 08 59.61 & 2.07 & 12.25 & 0.90*  \\
 S08  & 20 20 51.286 & +40 15 19.81 & (0.095, 0.155) & 0.37 & 15.45 & 0.023* & 20 20 51.274 & +40 15 19.69 & 0.29 & 14.86 & 0.008* \\
 S09  & 20 20 52.606 & +40 24 30.33 & (0.063, 0.064) & 2.47 & 15.39 & 1.02   & 20 20 52.608 & +40 24 30.07 & 2.57 & 12.34 & 0.71*  \\
 S10  &              &              &                &      &       &        & 20 20 54.954 & +40 24 19.04 & 0.82 & 14.34 & 0.34*  \\
 S11  & 20 20 57.879 & +40 10 50.13 & (0.112, 0.051) & 0.23 & 16.45 & 0.009* & 20 20 57.890 & +40 10 50.03 & 0.14 & 13.70 & 0.007* \\
 S13  & 20 21 11.802 & +40 10 40.81 & (0.055, 0.088) & 0.86 & 15.08 & 0.12*  & 20 21 11.794 & +40 10 40.56 & 0.95 & 13.02 & 0.17*  \\
 S14  & 20 21 12.918 & +40 24 05.67 & (0.024, 0.174) & 0.34 & 17.71 & 0.038* & 20 21 12.912 & +40 24 05.54 & 0.24 & 13.51 & 0.016* \\
 S16  & 20 21 20.374 & +40 17 28.49 & (0.125, 0.305) & 0.53 & 14.60 & 0.041* & 20 21 20.378 & +40 17 27.96 & 0.25 & 9.66  & 0.001* \\
 S21  & 20 21 30.885 & +40 26 47.96 & (0.494, 0.590) & 3.06 & $>19$  & 1.32   &              &              &      &       &        \\
 S23  & 20 21 30.825 & +40 11 22.98 & (0.031, 0.061) & 3.95 & 13.62 & 1.82   & 20 21 30.828 & +40 11 22.62 & 3.64 & 12.20 & 1.20   \\
 S24  & 20 21 31.425 & +40 22 56.71 & (0.164, 0.094) & 0.92 & 17.47 & 0.27*  & 20 21 31.422 & +40 22 56.41 & 0.92 & 13.19 & 0.18*  \\
 S26  & 20 21 33.166 & +40 15 57.06 & (0.036, 0.066) & 0.63 & 14.53 & 0.056* & 20 21 33.173 & +40 15 56.89 & 0.81 & 11.39 & 0.03*  \\
 S28  &              &              &                &      &       &        & 20 21 34.983 & +40 23 58.99 & 0.89 & 15.02 & 0.65*  \\
 S29  & 20 21 46.392 & +40 17 59.56 & (0.047, 0.196) & 1.36 & 13.83 & 0.23*  & 20 21 46.302 & +40 17 59.91 & 1.75 & 12.24 & 0.30*  \\
\enddata \tablecomments{\\
$^a$ USNO RMS positional uncertainty in each axis (RA, Dec)\\
$^b$ 2MASS RMS positional uncertainty $\sigma_{i}=0\farcs08$ per axis (http://www.ipac.caltech.edu/2mass/releases/allsky/doc/sec6$\_6$a.html)\\
$*$  A strong-candidate counterpart---$\delta < \epsilon_{99}$ and $p(\delta, {\rm I|K}) < 1\%$---as discussed in the text}
 \end{deluxetable}

\clearpage

\begin{deluxetable}{ccccccccccccccc}
\tabletypesize{\scriptsize}
\tablewidth{0pc}
\tablecaption{2MASS infrared photometry of strong-candidate counterparts to X-ray sources. \label{t:2MASS_colors}}
\tablehead{(1) & (2) & (3) & (4) & (5) & (6) & (7) & (8) & (9)}
\startdata
Source & J & H & K$_s$ & J$-$H & H$-$K$_s$ & J$-$K$_s$ & $S_{\nu}$\,@\,139\,THz & $\alpha_{ix}$ \\
       &   &   &       &       &           &           & mJy & \\ \hline\\[-2ex]
S01 & 14.154$\pm$0.029 & 13.434$\pm$0.031 & 13.329$\pm$0.041 & 0.072$\pm$0.042 & 0.105$\pm$0.051 & 0.825$\pm$0.050 &  3.11$\pm$0.13 & 1.86 \\
S02 & 12.608$\pm$0.024 & 12.040$\pm$0.020 & 11.884$\pm$0.018 & 0.568$\pm$0.031 & 0.156$\pm$0.027 & 0.724$\pm$0.030 & 11.76$\pm$0.21 & 2.12 \\
S04 & 12.425$\pm$0.025 & 11.944$\pm$0.021 & 11.823$\pm$0.020 & 0.481$\pm$0.033 & 0.121$\pm$0.029 & 0.602$\pm$0.032 & 12.44$\pm$0.25 & 2.13 \\
S05 & 12.862$\pm$0.026 & 12.241$\pm$0.021 & 12.048$\pm$0.022 & 0.621$\pm$0.033 & 0.193$\pm$0.030 & 0.814$\pm$0.034 & 10.11$\pm$0.22 & 2.01 \\
S06 & 16.358$\pm$0.121 & 15.058$\pm$0.099 & 14.653$\pm$0.124 & 1.300$\pm$0.156 & 0.405$\pm$0.159 & 1.705$\pm$0.173 &  0.92$\pm$0.11 & 1.80 \\
S07 & 13.365$\pm$0.026 & 12.592$\pm$0.021 & 12.246$\pm$0.022 & 0.773$\pm$0.033 & 0.346$\pm$0.030 & 1.119$\pm$0.034 &  3.56$\pm$0.14 & 1.97 \\
S08 & 16.089$\pm$????  & 15.216$\pm$0.095 & 14.855$\pm$0.143 & 0.873$\pm$0.095 & 0.361$\pm$0.172 & 1.234$\pm$0.143 &  8.43$\pm$0.19 & 2.13 \\
S09 & 13.503$\pm$0.026 & 12.695$\pm$0.021 & 12.338$\pm$0.030 & 0.808$\pm$0.033 & 0.357$\pm$0.037 & 1.165$\pm$0.040 &  7.74$\pm$0.23 & 1.94 \\
S10 & 15.903$\pm$0.087 & 14.892$\pm$0.087 & 14.343$\pm$0.095 & 1.011$\pm$0.123 & 0.549$\pm$0.129 & 1.560$\pm$0.129 &  1.22$\pm$0.12 & 1.56 \\
S11 & 14.819$\pm$0.040 & 14.054$\pm$0.040 & 13.698$\pm$0.053 & 0.765$\pm$0.057 & 0.356$\pm$0.066 & 1.121$\pm$0.066 &  2.21$\pm$0.12 & 1.91 \\
S13 & 13.843$\pm$0.023 & 13.217$\pm$0.024 & 13.023$\pm$0.029 & 0.626$\pm$0.033 & 0.194$\pm$0.038 & 0.820$\pm$0.037 &  4.12$\pm$0.12 & 1.98 \\
S14 & 15.010$\pm$0.044 & 13.877$\pm$0.036 & 13.508$\pm$0.048 & 1.133$\pm$0.057 & 0.369$\pm$0.060 & 1.502$\pm$0.065 &  2.64$\pm$0.13 & 1.84 \\
S16 & 11.388$\pm$0.021 & 10.135$\pm$0.018 &  9.661$\pm$0.015 & 1.253$\pm$0.028 & 0.474$\pm$0.023 & 1.727$\pm$0.026 & 91.12$\pm$1.37 & 2.23 \\
S24 & 14.761$\pm$0.031 & 13.523$\pm$0.027 & 13.188$\pm$0.035 & 1.238$\pm$0.041 & 0.335$\pm$0.044 & 1.573$\pm$0.047 &  3.54$\pm$0.12 & 1.97 \\
S26 & 12.597$\pm$0.021 & 11.713$\pm$0.019 & 11.392$\pm$0.018 & 0.884$\pm$0.028 & 0.321$\pm$0.026 & 1.205$\pm$0.028 & 18.50$\pm$0.33 & 2.03 \\
S28 & 16.650$\pm$0.148 & 15.595$\pm$0.122 & 15.017$\pm$0.167 & 1.055$\pm$0.192 & 0.578$\pm$0.207 & 1.633$\pm$0.223 &  0.66$\pm$0.11 & 1.62 \\
S29 & 12.761$\pm$0.022 & 12.332$\pm$0.019 & 12.239$\pm$0.023 & 0.429$\pm$0.029 & 0.093$\pm$0.030 & 0.522$\pm$0.032 &  8.58$\pm$0.20 & 2.06
 \enddata
 \end{deluxetable}

\clearpage

\begin{deluxetable}{llll}
\tabletypesize{\scriptsize}
\tablewidth{0pc}
\tablecaption{Log of optical observations of targets in the \EG\ field.\label{t:obs_log}}
\tablehead{\multicolumn{1}{c}{(1)} & \multicolumn{1}{c}{(2)} & \multicolumn{1}{c}{(3)} & \multicolumn{1}{c}{(4)}}
\startdata
Date   & Instrument (Conditions) & Objects & Observation type \\
2005   & &  &  \\ \hline\\[-2ex]
May 30 & LFOSC           & S03; S03+S02+S05+S08            & Deep RI imaging \\
       & (intermittent or thin clouds) & S01+S06+S07; S02+S03+S04+S05; & Short R imaging \\
       &  & S04+S02+S05+S09+S10; S05+S02+S03+S04+S12 & Short R imaging \\
       &                 & S01; S02; S04; S05 & Spectroscopy \\
May 31 & LFOSC & S06+S07+S11+S13+S18+S23+S27  & RI imaging \\
       & (mediocre weather) & S08 & Spectroscopy \\
June 1 & None (bad weather) &     & None \\
June 2 & Boller \& Chivens (clear) & S02; S04; S08; S23; S27; S29  & Spectroscopy \\
June 3 & Boller \& Chivens (clear) & S05; S09; S13; S24 & Spectroscopy \\
June 4 & C\'amara Directa (clear)  & S03; S06+S07+S11; S09+S10; S12; S13+S15; S14; & V imaging \\
       & (CCD warmed $\Rightarrow$ bad V images) & S21+S25; S22+S23; S28+S24+S25; S29+S30 & V imaging \\
       &   & S12; S13+S15; S14; S21+S25+S28; S22+S23;  & R imaging \\
       &   & S28+S24+S25, S29+S30 & R imaging \\
June 5 & C\'{a}mara Directa (clear)  & S03; S06+S07+S11; S09+S10; S16+S17 & R imaging \\
       &   & S03; S06+S07+S11; S09+S10; S12; S13+S15; S14; & I imaging\\
       &   & S20+S26; S21; S22+S23; S28+S24+S25; S29+S30; & I imaging\\
       &   & S16+S17 & I imaging \\
       & (bad V image) & S16  & V imaging \\
June 7 & Boller \& Chivens (clear) & S08; S09; S11; S16; S26 & Spectroscopy
 \enddata
 \end{deluxetable}

\clearpage
\begin{deluxetable}{ccccccll}
\tabletypesize{\scriptsize}
\rotate
\tablewidth{0pc}
\tablecaption{Optical imaging and spectroscopy of targets in the \EG\ field. \label{t:obs_data}}
\tablehead{(1) & \multicolumn{1}{c}{(2)} & \multicolumn{1}{c}{(3)} & \multicolumn{1}{c}{(4)} & \multicolumn{1}{c}{(5)} & \multicolumn{1}{c}{(6)} & \multicolumn{1}{c}{(7)} & \multicolumn{1}{c}{(8)}}
\startdata
 Source & Candidate & \multicolumn{4}{c}{Photometry}
 & Spectroscopy& Comments \\
 & counterpart & R & I & J & H & & (Chance, if $\ge\! 1\%$) \\ \hline\\[-2ex]
S01 & USNO, 2MASS & $\approx\!16.1$ & & & & Mg, Na, no Balmer & G or later \\
S02 & USNO, 2MASS & $14.8\!\pm\!0.1$ & & & & Mg, Na, no Balmer &  Late G \\
S03 & None & $>\!23.0$ & $>\!21.8$ & & $>\!20$ & & No candidate $<\!3\arcsec$ \\
S04 & USNO, 2MASS & $14.3\!\pm\!0.1$ & & & & Mg, Na, Fe, weak Balmer & Late A or early F \\
S05 & USNO, 2MASS & $15.6\!\pm\!0.1$ & & & & Mg, Na, Fe, weak Balmer & Late A or early F \\
S06 & 2MASS & $21.2\!\pm\!0.4$ & $17.8\!\pm\!0.1$ & $16.0\!\pm\!0.1$ & $14.9\!\pm\!0.1$ & &  \\
S07 & 2MASS & & $18.9\!\pm\!0.1$ & & & & \\
S08 & USNO, 2MASS & $17.5\!\pm\!0.3$ & & & & Mg?, weak Na, no Balmer & G or later? \\
S09 & (USNO), 2MASS & $17.0\!\pm\!0.1$ & $15.1\!\pm\!0.1$ & & & Low S/N & (1.0\%) \\
S10 & 2MASS & $>\!22.9$ & $18.3\!\pm\!0.1$ & & $14.8\!\pm\!0.1$ & & \\
S11 & USNO, 2MASS & $17.8\!\pm\!0.1$ & $16.1\!\pm\!0.1$ & & & Low S/N & \\
S12 & None & $>\!23.5$ & $>\!22.1$ & & $>\!19$ & & No candidate $<\!3\arcsec$ \\
S13 & USNO, 2MASS & $16.7\!\pm\!0.1$ & $15.2\!\pm\!0.1$ & & & Low S/N; Mg?, H$\beta$? & Late A or early F? \\
S14 & USNO, 2MASS & $19.1\!\pm\!0.1$ & $16.4\!\pm\!0.1$ & & & & \\
S15 & None & $>\!23.2$ & $>\!22.5$ & & & & No candidate $<\!3\arcsec$ \\
S16 & USNO, 2MASS & $16.8\!\pm\!0.1$ & $14.2\!\pm\!0.1$ & & & Low S/N & \\
S17 & None & $>\!19.5$ & $>\!18.8$ & & & & No candidate $<\!3\arcsec$ \\
S18 & None & $>\!20.0$ & $>\!18.0$ & & & & No candidate $<\!3\arcsec$ \\
S19 & Unobserved & & & & & & No observations \\
S20 & None & & $>\!22.0$ & & & & No candidate $<\!3\arcsec$ \\
S21 & (USNO) None & $>\!22.8$ & $>\!19.7$ & & & & (1.3\%) Not found $<\!3\arcsec$ \\
S22 & None & $>\!22.7$ & $>\!22.5$ & $>\!17.9$ & $>\!17.9$ & & No candidate $<\!3\arcsec$ \\
S23 & (USNO, 2MASS); (New) & $>\!22.7$ & $>\!21.6$ & $17.3\!\pm\!0.1$ & $16.9\!\pm\!0.3$ & & (1.8\%, 1.2\%); ($1\farcs2$ NW, $\approx\!$3.8\%) \\
S24 & USNO, 2MASS & $19.9\!\pm\!0.1$ & $16.9\!\pm\!0.1$ & & & & \\
S25 & (New) & $>\!23.5$ & $>\!22.1$ & & $17.3\!\pm\!0.1$ & & ($0\farcs8$ SE, $\approx\!2.3\%$)\\
S26 & USNO, 2MASS & $16.2\!\pm\!0.1$ & $17.2\!\pm\!0.1$ & & & Na, no Balmer & Late-type?\\
S27 & None & $\ga\!20.0$ & $\ga\!18.8$ & $>\!17.9$ & $>\!17.9$ & & No candidate $<\!3\arcsec$ \\
S28 & 2MASS & $22.1\!\pm\!0.8$ & $19.9\!\pm\!0.6$ & & $15.8\!\pm\!0.1$ & & \\
S29 & USNO, 2MASS & $14.3\!\pm\!0.1$ & $13.5\!\pm\!0.1$ & & & Mg, Na, H$\beta$ & A or F \\
S30 & None & $>\!23.2$ & & $>\!17.9$ & $>\!17.9$ & & No candidate $<\!3\arcsec$
\enddata
\end{deluxetable}

\begin{deluxetable}{cccccccc}
\tabletypesize{\scriptsize}
\tablewidth{0pc}
\tablecaption{WIRC observations of targets in the \EG\ field. \label{t:wirc_data}}
\tablehead{(1) & (2) & (3) & (4) & (5) & (6) & (7) & (8) }
\startdata
Source & RA(J2000)   & Dec(J2000)  & $\delta_{ix}$ & K$_s$ & $p_{i}(\delta_{ix},{\rm K}_{s})$ & $S_{\nu}$\,@\,139\,THz & $\alpha_{ix}$ \\
& $^{\rm h}\: \ ^{\rm m}\ \ ^{\rm s}\: $\ \ \ \ & \ $\ \arcdeg\: \ \ \arcmin\: \ \ \arcsec$\ \ \ & $\arcsec$ & & \% & mJy &   \\ \hline\\[-2ex]
S03 & 20 20 33.73 & 40 18 28.2 &  0.52 & 18.79$\pm$0.20 & 1.61  & 0.002 $\pm$ 0.004 & 1.13 \\
S12 & 20 21 04.13 & 40 21 39.5 &  0.13 & 17.03$\pm$0.06 & 0.07* & 0.103 $\pm$ 0.006 & 1.30 \\
S17 & 20 21 23.78 & 40 19 03.4 &  0.22 & 15.33$\pm$0.02 & 0.06* & 0.492 $\pm$ 0.010 & 1.72 \\
S19 & 20 21 26.38 & 40 20 09.0 &  2.25 & 17.30$\pm$0.07 & 24.6  & 0.080 $\pm$ 0.006 & 1.50 \\
S20 & 20 21 26.68 & 40 14 52.2 &  0.33 & 19.19$\pm$0.20 & 0.70* & 0.014 $\pm$ 0.003 & 1.22 \\S25 & 20 21 32.54 & 40 24 59.6 &  1.23 & 16.89$\pm$0.04 & 5.88  & 0.117 $\pm$ 0.005 & 1.26 \\
S30 & 20 21 47.25 & 40 18 49.2 &  1.61 & 18.23$\pm$0.10 & 16.3  & 0.034 $\pm$ 0.003 & 1.31
 \enddata
 \end{deluxetable}

\clearpage

\clearpage

\clearpage
\begin{figure}
\begin{center}
\epsfig{file=f3a.eps,width=4.cm,angle=-90}
\epsfig{file=f3b.eps,width=4.cm,angle=-90}
\epsfig{file=f3c.eps,width=4.cm,angle=-90}
\epsfig{file=f3d.eps,width=4.cm,angle=-90}
\epsfig{file=f3e.eps,width=4.cm,angle=-90}
\epsfig{file=f3f.eps,width=4.cm,angle=-90}
\epsfig{file=f3g.eps,width=4.cm,angle=-90}
\epsfig{file=f3h.eps,width=4.cm,angle=-90}
\epsfig{file=f3i.eps,width=4.cm,angle=-90}
\epsfig{file=f3j.eps,width=4.cm,angle=-90}
\epsfig{file=f3k.eps,width=4.cm,angle=-90}
\epsfig{file=f3l.eps,width=4.cm,angle=-90}
\epsfig{file=f3m.eps,width=4.cm,angle=-90}
\epsfig{file=f3n.eps,width=4.cm,angle=-90}
\epsfig{file=f3o.eps,width=4.cm,angle=-90}
\end{center}
\figcaption{X-ray spectra for sources S16--S30, showing count rate per 0.234-keV bin versus energy (in keV).
Arranged left-to-right, S16 is at the upper left and S30, lower right.
\label{f:spectra_2}}
\end{figure}

\clearpage

\clearpage

\clearpage

\clearpage

\clearpage

\end{document}